# A Controllable Appearance Representation for Flexible Transfer and Editing

Santiago Jimenez-Navarro, Julia Guerrero-Viu & Belen Masia

Universidad de Zaragoza, I3A, Spain

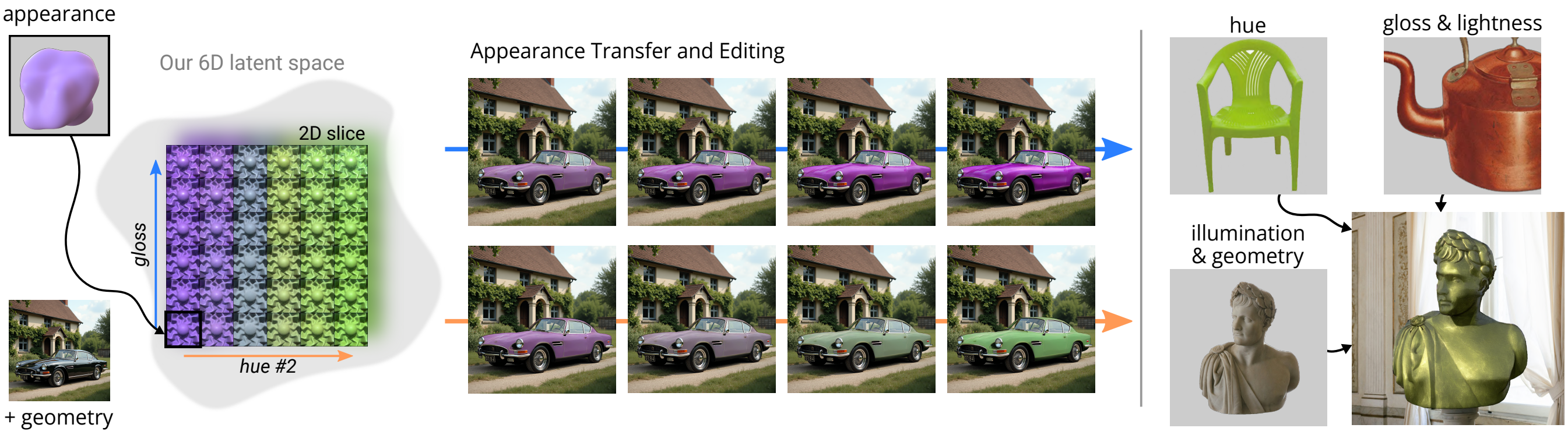

**Figure 1:** *Our model learns a disentangled and interpretable latent space of appearance in a self-supervised manner, without human-annotated data. Given an input image depicting a homogeneous object (left), it can be encoded into this space, which can then be traversed to generate meaningful variations of appearance (here, traversals along two dimensions encoding hue and gloss are shown). This encoded representation of appearance can be leveraged for appearance transfer and editing tasks: Given a target geometry (bottom left), we can transfer to it either the original appearance, or variations along any of the dimensions of the space (center). Since the space is disentangled and interpretable, it also enables selective appearance transfer (right): The resulting image (bottom right) is generated by selecting specific dimensions from each of the three inputs.*

**Abstract**

*We present a method that computes an interpretable representation of material appearance within a highly compact, disentangled latent space. This representation is learned in a self-supervised fashion using a VAE-based model. We train our model with a carefully designed unlabeled dataset, avoiding possible biases induced by human-generated labels. Our model demonstrates strong disentanglement and interpretability by effectively encoding material appearance and illumination, despite the absence of explicit supervision. To showcase the capabilities of such a representation, we leverage it for two proof-of-concept applications: image-based appearance transfer and editing. Our representation is used to condition a diffusion pipeline that transfers the appearance of one or more images onto a target geometry, and allows the user to further edit the resulting appearance. This approach offers fine-grained control over the generated results: thanks to the well-structured compact latent space, users can intuitively manipulate attributes such as hue or glossiness in image space to achieve the desired final appearance.*

## 1. Introduction

The visual appearance of a material in an image arises from the complex interplay of the material's reflectance properties, the lighting conditions, and the geometry of the object it is applied to. The combination of these gives rise to the proximal stimulus, which reaches our retinae and is interpreted by our visual system. Ade-quately characterizing material appearance is a fundamental goal of computer graphics. Traditionally, material properties are modeled through reflectance distribution functions, expressed either analytically, or in the form of tabulated data [MPBM03; DJ18]. More recent works rely on neural processes [ZZW*21] or SVBRDF maps, which allow characterization of complex textured materials from a single or few images [DAD*18; VMR*24]. These representations, while well suited for rendering, suffer from high-dimensionality, limited expressiveness, or a lack of interpretability, which hinders downstream tasks such as material compression or editing.

For these reasons, a number of works have been devoted to finding compact representations of material appearance, focusing on compression [HGC*20], interpolation [SSN18], editability [SWSR21], or the perceptual nature of the representation [SGM*16]. Depending on the application domain, different properties, such as interpretability or independence of dimensions, may be desirable in such representation. When aiming for interpretable spaces, existing approaches often rely on large quantities of human-annotated data. These are very costly to obtain, even more so across different attributes [SCW*21; DLC*22]. Besides, it is unclear a priori which attributes these should be [MPBM03; SGM*16; TGG*20]. Consequently, in this work we explore the use of self-supervised learning for identifying the underlying factors that determine material appearance, avoiding the need for labeled data in the creation of an interpretable and controllable latent space.

We leverage FactorVAE [KM18], a well-known method for disentangled representation learning, and build upon it to adapt it to our scenario. Specifically, we modify its original architecture to incorporate geometry information in the decoder, enforcing the bottleneck to learn the appearance separate from the geometry, and we also modify its loss function to avoid posterior collapse in our setup. Further, we carefully design a synthetic dataset that enables the FactorVAE to learn, in a self-supervised manner, explainable and independent dimensions encoding material appearance of homogeneous, opaque objects. Notably, and unlike many previous approaches, we work in image space, which has two advantages: first, we can encode the appearance of objects from images in the wild, and second, our model encodes visual appearance of the material, and not only material properties.

Our appearance encoder model thus enables encoding an input image, depicting an object, into a six-dimensional disentangled representation of the appearance of the object in the image. Despite the model being trained without labeled data, the six dimensions are interpretable, and encode hue (two dimensions), illumination (two dimensions), lightness and glossiness (note that both the illumination and the reflectance properties are included in this appearance representation). Since they are disentangled, the latent variable controlling gloss will change only with variation in object gloss, while the rest will remain constant. This greatly facilitates controllability, and therefore applications like editing, or selective attribute transfer. Fig. 1, left, shows the traversal of our learned latent space along the subspace spanned by two of its dimensions.

We demonstrate the potential of this disentangled, interpretable and controllable space for two applications: appearance transfer and editing (Fig. 1, center). We do this by using the latent appearance representation as a guidance to train a second model: a lightweight IP-Adapter [YZL*23] which effectively translates the information from the latent space to a diffusion pipeline, leveraging their generative ability. Specifically, we use a pre-trained latent diffusion model, based on Stable Diffusion XL [PEL*], and condition the generative process through two distinct branches: one that encodes the appearance via our latent space, and another one to integrate the target geometry. In this way, the appearance branch can either encode directly the appearance of a material in an input image (transfer) or be edited to generate variations of an existing material (editing). Interestingly, under this scheme, appearance transfer can be done from a single input image, or from multiple, performing selective attribute transfer from each; an example of this is shown in Fig. 1, right, where the appearance in the final image results from combining the hue of an exemplar, the gloss and lightness of another, and the geometry and illumination of a third one.

Performing appearance transfer in this way has advantages over existing approaches [CSM*24], since it offers a better disentanglement between appearance and geometry, and therefore more control over the transfer. This increased control is also an advantage of our method when compared to other methods for diffusion-based image editing [BHE23]: Diffusion models are typically trained on text-image pairs, relying on text prompts as the primary conditioning method. Despite its wide expressivity, the *text-only* control introduces ambiguities that can significantly limit its application for the particular case of material appearance, thus benefiting from image-based conditioning.

Overall, our latent representation offers increased controllability and interpretability, demonstrating its potential for appearance transfer and editing. We therefore make the following contributions:

- A self-supervised model that encodes an object in an input image into a disentangled and interpretable representation of its appearance.
- A large-scale dataset of almost 100,000 synthetic images carefully designed for self-supervised learning of homogeneous material appearance.
- Application of our representation to flexible appearance transfer and editing, through conditioning of a diffusion-based pipeline.

Our trained models, code, and dataset are available in `http://graphics.unizar.es/projects/mat_disentanglement_2025/`

## 2. Related Work

### 2.1. Low-Dimensional Material Appearance Representations

Material appearance is shaped by complex interactions between several factors, including surface properties, lighting, geometry, or viewing conditions [FDA03; LSGM21], often requiring high-dimensional data for accurate modeling. A number of works have attempted to find a compact representation of appearance, searching for low-dimensional BRDF or BTF embeddings [MPBM03; RJGW19; RGJW20; Kuz21], enabling applications such as compression or editing. However, they usually require costly human-annotated datasets [SGM*16; TGG*20; SWSR21], or produce spaces that lack interpretability because their focus is on some other aspect, such as compression [HGC*20; SSN18; ZZW*21]. While the former work in material space, a series of approaches have focused on working directly in image space, to account for the interplay of confounding factors like geometry or illumination in the final perception of appearance [LMS*19; SCW*21]. Still, they rely on supervised learning-based methods and require large amounts of human annotations, which can be partially alleviated with weak supervision [GSS*24]. Some methods have focused directly on material editing in image space [DLC*22; SL23], leveraging GAN-based frameworks to allow controlled modification of specific attributes, like glossiness or metallicness, and also requiring ground-truth human labels for training. Unsupervised learning of appearance representations has been mostly limited to specific attributes, like gloss [SAF21] or translucency [LSX23], training on datasets with limited variability. In this work, we present a self-supervised model that learns a highly-compact latent space of material appearance. Remarkably, our model can disentangle interpretable factors like color or glossiness in images depicting homogeneous real-world materials, without any prior knowledge or annotated data.

### 2.2. Disentangled Representation Learning

Disentangled representation learning [WCWZ*24; LBL*19] aims to separate underlying factors of variation within data, improving interpretability in generative models. Variational Autoencoders (VAEs), including βVAE [HMP*17], achieve this by balancing reconstruction fidelity and latent space regularization, while FactorVAE [KM18] introduces a Total Correlation (TC) penalty via a discriminator network to enhance factor independence. Extensions like βTCVAE [CLGD18] propose different implementations of this TC term. A key challenge in these models is the posterior collapse, where latent variables lose informativeness by matching the prior too closely. Due to its relevance, this issue has been widely addressed [SRM*16; FLL*19; YWY*20; KOF*23]. Beyond VAEs, GAN-based [CDH*16] and diffusion models [YWLZ23] have also been explored for disentanglement. Nevertheless, the explicit modeling of a latent space in VAE-based models facilitates learning independent factors within a compact representation. Closer to our work is that of Benamira et al. [BSP22], that used βVAE to disentangle material appearance from measured BRDFs. In contrast, our model disentangles material appearance in *image space*, accounting for the influence of factors like illumination or geometry. We adapt FactorVAE to mitigate posterior collapse, and enable a self-supervised disentangled latent space to encode and modify material appearance (see Sec. 3).

### 2.3. Diffusion-Based Material Transfer and Editing

Since the seminal work from Ho et al. [HJA20], diffusion models have revolutionized image generation with exceptional quality and diversity, by progressively denoising from random noise [SCS*22; BGJ*23; RBL*22]. Conditioning mechanisms have recently played a central role to expand the functionality of diffusion models for *controlled* generation and editing. These include techniques like ControlNet [ZRA23] for multi-modal conditioning, LoRA [HSW*22] for task-specific fine-tuning, or IP-Adapter [YZL*23] and T2I-Adapter [MWX*24] for image-based conditioning, by injecting features from reference images alongside text prompts to influence the generation.

In the context of material appearance, diffusion models have been used for material generation [ZLX*24], physically-based synthesis [VBP*24], material capture [VMR*24], or texture editing [GHR*24]. In ColorPeel [BWVvdW24], they condition diffusion models on disentangled properties like color and texture, allowing for fine-grained editing but requiring labeled data to train the conditioning, which complicates generalization to novel properties. Cheng et al. [CSM*24] demonstrated zero-shot material transfer by injecting CLIP [RKH*21] embeddings of reference materials into a diffusion pipeline, achieving compelling results without any additional model training. However, their reliance on CLIP limits disentanglement between material properties and additional factors, such as lighting and geometry. Alchemist [SJL*24] showed impressive performance on material editing in image space, training diffusion models for specific material attributes with supervised ground-truth labels.

In our work, we train an IP-Adapter to condition a pre-trained diffusion model based on Stable Diffusion-XL [PEL*] on our compact self-supervised appearance representation, to showcase proof-of-concept applications of it (namely, appearance transfer and editing).

## 3. A Disentangled Latent Space for Appearance

In this section, we seek a material latent representation that disentangles the underlying factors responsible for its appearance in a given image. To do this, we train a variational autoencoder that reconstructs the appearance of an input image, while enforcing a latent space with independent dimensions encoding this appearance. The model (Sec. 3.1) takes as input an image of an object, made from a homogeneous material, and encodes it into a disentangled latent representation of the material's appearance in that image; this representation includes both the reflectance and the illumination. It can also decode such a representation, together with an input normal map, into an image of an object made of such material, whose geometry is determined by the input normals. This model is trained in a self-supervised manner, leveraging a dataset of almost 100,000 images specifically created for representation learning of material appearance (Sec. 3.2). We analyze the resulting latent space and the model's reconstruction ability in Sec. 4.

### 3.1. Method

We employ an encoder-decoder architecture to create a bottleneck of reduced dimensionality that encapsulates the internal representation of appearance (Fig. 2). We build our model on FactorVAE [KM18], a variational autoencoder for self-supervised learning of disentangled representations, for its effective balance between disentanglement and reconstruction quality. We introduce two key modifications to the original FactorVAE to make it suitable for our goal: (1) we modify the loss to avoid posterior collapse (Sec. 3.1.1), and (2) we input geometry information to the model in the form of normal maps, compelling the latent space to focus on appearance (Sec. 3.1.2).

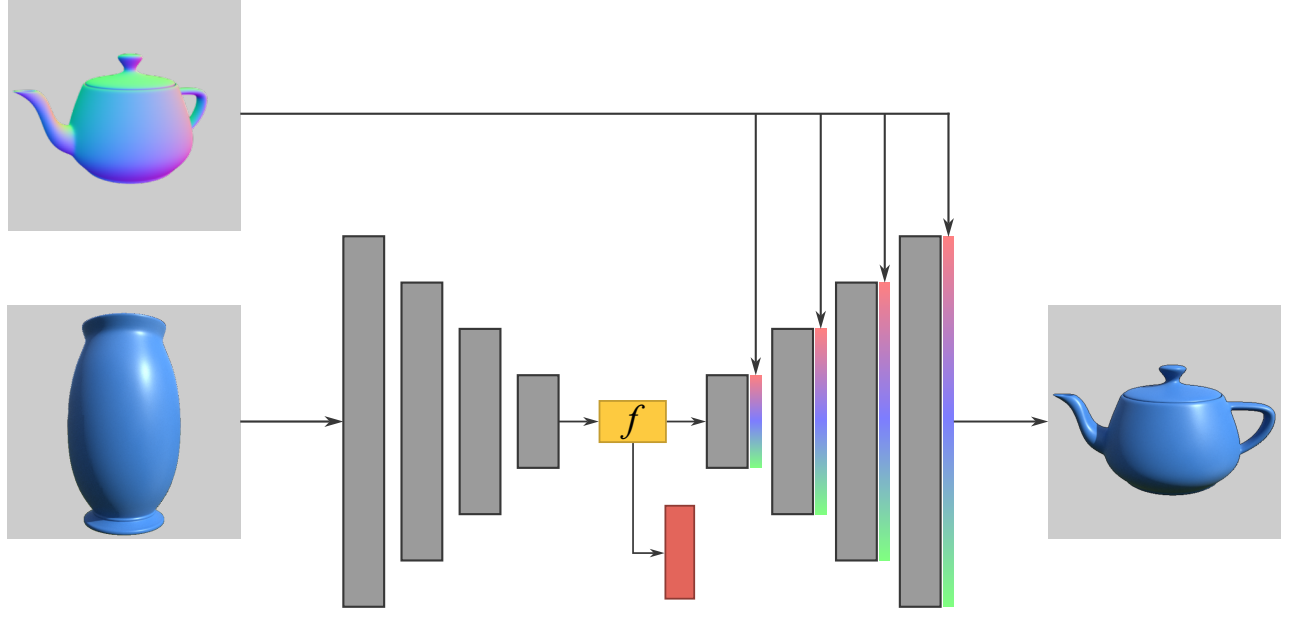

**Figure 2:** ***Diagram of our VAE-based architecture.*** *The encoder creates a low-dimensional, disentangled representation f of the appearance of the input image. The decoder learns to apply the incoming appearance to a reference geometry, specified with a normal map, which is concatenated in the decoder pipeline. The red box illustrates the discriminator used to compute the TC term (see text for details).*

### 3.1.1. Avoiding Posterior Collapse

The training loss proposed in the FactorVAE paper enforces the reconstruction of the input image, and the informativeness and independence of the dimensions of the latent space. To do so, it has three distinct terms: a term for reconstruction quality, a regularization term, and a term enforcing independence between factors (or Total Correlation term, TC). VAE-based models like this one, however, often suffer from a phenomenon called *posterior collapse*, in which the distribution learned by the encoder collapses to its prior, thus storing no relevant information about the generative factors of data [KM18; LTGN19]. Our proposed loss function $\mathcal{L}_{\theta,\phi}$ therefore is based on that of FactorVAE, with some modifications to avoid posterior collapse. It has the following formulation:

$$\mathcal{L}_{\theta,\phi}(x) = \mathbb{E}_{z\sim q_\theta(z|x)}\left[\log p_\phi(x \mid z)\right] - \beta D_{KL}(q_\theta(z\mid x), p(z), n) - \gamma D_{KL}(q_\theta(z), \bar{q}_\theta(z), 1), \tag{1}$$

where

$$\bar{q}_\theta(z) := \prod_{j=1}^{d} q_\theta\left(z_j\right), D_{KL}(Q,P,n) := \left\| \frac{1}{2}(\mu^2+\sigma^2 - log(\sigma^2) - 1)\right\|^n$$

In Eq. 1, the probabilistic encoder $p(z \mid x)$ is approximated to the distribution $q_\theta(z \mid x)$, where $\theta$ corresponds to the weights learned by the network, and $x$ describes a sample represented as $z$ in the latent space. Analogously, $p_\phi(x \mid z)$ represents the probabilistic decoder, approximated with the weights $\phi$.

For the reconstruction term we use a smooth L1 loss [Gir15] between the input image and the reconstructed one.

The second term serves as a regularizer of the latent space by minimizing the dimension-wise Kullback-Leibler (KL) divergence between the learned distribution and a prior $p(z)$. Being a variational model, the standard choice for the prior is a normal distribution. This term contains two key modifications with respect to the original formulation, aimed to address the posterior collapse issue. First, we apply a norm of order $n$ (instead of the default summation) to the result of the KL operator. This encourages all dimensions to have a similar distance to the prior, thus storing a similar amount of information, effectively countering the posterior collapse. Second, we extend the FactorVAE loss by incorporating a weight $\beta$ in this term, not present in the original definition. This weight allows, during training, to specify how much the learned distributions should resemble the prior. Additionally, we apply a linear annealing to the $\beta$ term during training [SRM*16]. Gradually increasing the adherence to the prior distribution encourages the model to distribute information more evenly across the latent dimensions in the early training stages, further mitigating posterior collapse.

The TC term [Wat60] encourages the model to learn *independent* latent dimensions. In our application, this translates to learning different attributes in each dimension of the latent space. Following the original implementation of FactorVAE [KM18], we compute this TC term using an external discriminator.

### 3.1.2. Enforcing Appearance Encoding

In order for our representation to focus on the appearance of the surfaces, and not on their geometry, we input geometrical information to the model, in the form of normal maps, in the decoder [DLC*22], as shown in Fig. 2. This encourages the representation learnt by the encoder to focus on the reflectance and illumination. As a result, the geometry will not be identified as a factor of variation in the latent space. Moreover, this approach significantly improves the model's generalization ability, enabling it to more efficiently learn the desired explainable and independent factors during training.

As a result of the aforementioned adaptations (which we ablate in Sec. 4.4 and the supplementary material (S4)), our model learns a latent space with six dimensions, where each dimension represents an independent factor of material appearance. This represents a substantial reduction of dimensionality, from an RGB image (with a spatial resolution of 256 × 256 in our implementation) into a 6D feature vector $f$ that successfully encodes the material's appearance, as shown in Sec. 4. Implementation details can also be found in the supplementary material (S1.1).

## 3.2. A Dataset for Material Appearance Disentanglement

Our model is trained in a self-supervised manner, learning to apply an input appearance onto a reference geometry, while building a well-structured latent space by optimizing the loss function (Eq. 1).

Existing datasets of material appearance are relatively abundant [SAF21; DLC*22; SCW*21]. However, they often include simple and unrealistic geometries with limited diversity, or are highly unbalanced with respect to appearance factors such as glossiness or hue. This is particularly problematic for our *self-supervised* training, as it could introduce unintended biases into the learned appearance representations. Moreover, we seek for a larger-scale dataset to improve generalization.

Therefore, we carefully designed a training dataset comprised of 98,550 synthetic images of 30 different objects rendered with 365 measured BRDFs [MPBM03; DJ18; SCW*21]. In order to facilitate the understanding of illumination by the network, we systematically vary the lighting for each object and material combination, leading to 9 lighting conditions (9 × 30 × 365 = 98,550).

**Table 1:** ***Quantitative metrics*** *obtained by our model and three baselines. All models are trained on our novel dataset (Sec. 3.2), and metrics are computed on our test dataset. Disentanglement and interpretability metrics are computed analyzing the structure of the latent space, using ground truth labels when necessary. Reported values are the mean and standard deviation of 5 independent trials, where each trial computes the metrics with a representative set of images. Arrows indicate desired performance, and the best value for each column is marked in bold.*

| | Disentanglement | | Interpretability | | Reconstruction quality | | |
|---|---|---|---|---|---|---|---|
| | GTC ↓ | MIS ↓ | Z-min ↑ | MIR ↑ | SSIM ↑ | LPIPS ↓ | PSNR ↑ |
| βVAE | - | 0.8825 ± 0.0197 | 0.5498 ± 0.0119 | 0.2224 ± 0.0046 | 0.5529 ± 0.0922 | 0.6642 ± 0.1638 | 29.129 ± 0.748 |
| βTCVAE | - | 0.8620 ± 0.0248 | 0.5717 ± 0.0061 | 0.2634 ± 0.0078 | 0.6177 ± 0.0916 | 0.6266 ± 0.1443 | 30.082 ± 1.490 |
| FactorVAE | 1.0633 | 0.8556 ± 0.0365 | 0.5521 ± 0.0041 | 0.2626 ± 0.0097 | 0.6556 ± 0.0596 | 0.5934 ± 0.0984 | 30.366 ± 1.290 |
| *Ours* | **0.6686** | **0.8204 ± 0.0514** | **0.6955 ± 0.0185** | **0.2940 ± 0.0089** | **0.6775 ± 0.0646** | **0.4046 ± 0.0648** | **30.801 ± 1.130** |

A representative set of images of the dataset, rendered with Mitsuba [Jak10], can be found in the supplementary material (S2).

## 4. Evaluation

We evaluate the ability of our model to find a disentangled latent space of material appearance, and to encode an input image into a suitable representation of the depicted material in this space. The evaluation is done both qualitatively and quantitatively, comparing its performance with various baselines. Specifically, we look at disentanglement and interpretability of the space, and at reconstruction quality of the model.

### 4.1. Quantitative Evaluation

Quantitatively, we compare our method to three baselines that are commonly used for disentangled representation learning (Sec. 2.2), all trained on our dataset (Sec. 3.2): βVAE [HMP*17], βTCVAE [CLGD18] and the vanilla FactorVAE [KM18]. The comparison evaluates disentanglement, interpretability and reconstruction quality. Our test set is comprised of images from the Serrano dataset [SCW*21]. From it, we select the ones featuring simple geometries (*blob* and *sphere*) to evaluate disentanglement and interpretability, to ensure that the lack of specialization of the baseline models in terms of geometric disentanglement is not overly penalized. To evaluate reconstruction quality we select images from a complex, unseen geometry (*statuette*). Note that, while some materials are both in our training and test sets, they are rendered with different illuminations and scene configurations, resulting in different appearance; besides, all baselines we compare to are trained and tested on the same sets.

Disentanglement is typically measured quantitatively with Gaussian Total Correlation (GTC) [Wat60] and the Mutual Information Score (MIS) [Sha48], which analyze the structure of the space and do not require labeled data. Measuring interpretability does require labeled data, and we measure it with the Z-min [KM18] and Mutual Information Ratio (MIR) [WDGB23] metrics, using ground truth labels available in the test dataset. We show all four metrics in Table 1. Our model clearly surpasses the rest, including the vanilla FactorVAE, showing the benefits of our modifications described in Sec. 3.1. Reconstruction quality is evaluated on the same test data with three widely-used metrics: SSIM [WBSS04], LPIPS [ZIE*18], and PSNR. Our model achieves better reconstructions, which can be attributed to the inclusion of geometry information in the decoder pipeline.

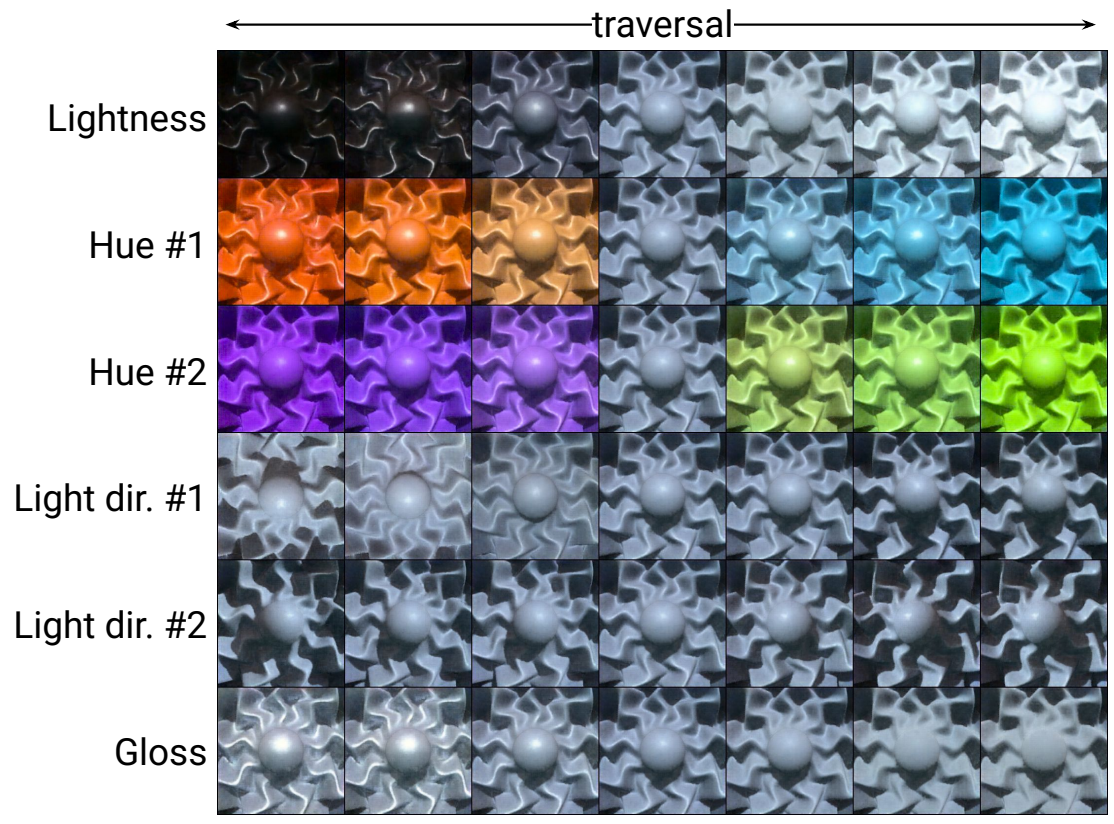

**Figure 3:** ***Prior traversals*** *sampling our 6D latent space. Each row samples a different dimension of the space, starting from a neutral, zero-valued feature vector (central column). The feature vector is fed to the decoder together with the normal map of the* Havran *geometry to generate the images shown. Our unsupervised model yields dimensions that are not only disentangled, but also interpretable, as indicated by the attributes we identify a posteriori.*

### 4.2. Qualitative Evaluation

Qualitatively, we can evaluate the disentanglement and interpretability of our latent space by visualizing the reconstructions generated by our decoder when sampling the latent space. To visualize the information encoded in the space, we take samples along each dimension (keeping the rest at a constant value of zero), generating feature vectors $f \in \mathbb{R}^6$ that are fed into the decoder together with a normal map. The results are the images shown in the *prior traversals* plot in Fig. 3. We chose the normal map of the *Havran* geometry for this visualization, as it is commonly used in perceptual studies, since it was specifically designed for broad light direction coverage, better showcasing a BRDF's appearance [HFM16]. In the figure, each row corresponds to the traversal of one latent dimension. We can see how the model identifies a slightly glossy gray material as the *neutral* one (zero-valued feature vector, central column). We can also clearly observe how, despite the lack of supervision, our model identifies interpretable factors that emerge from the data: Dimension 1 encodes lightness, dimensions 2 and 3 correspond to hue, dimensions 4 and 5 encode lighting direc-tion (top to bottom and left to right, respectively), and dimension 6 represents gloss. Additional visualizations can be found in the supplemental material (S3).

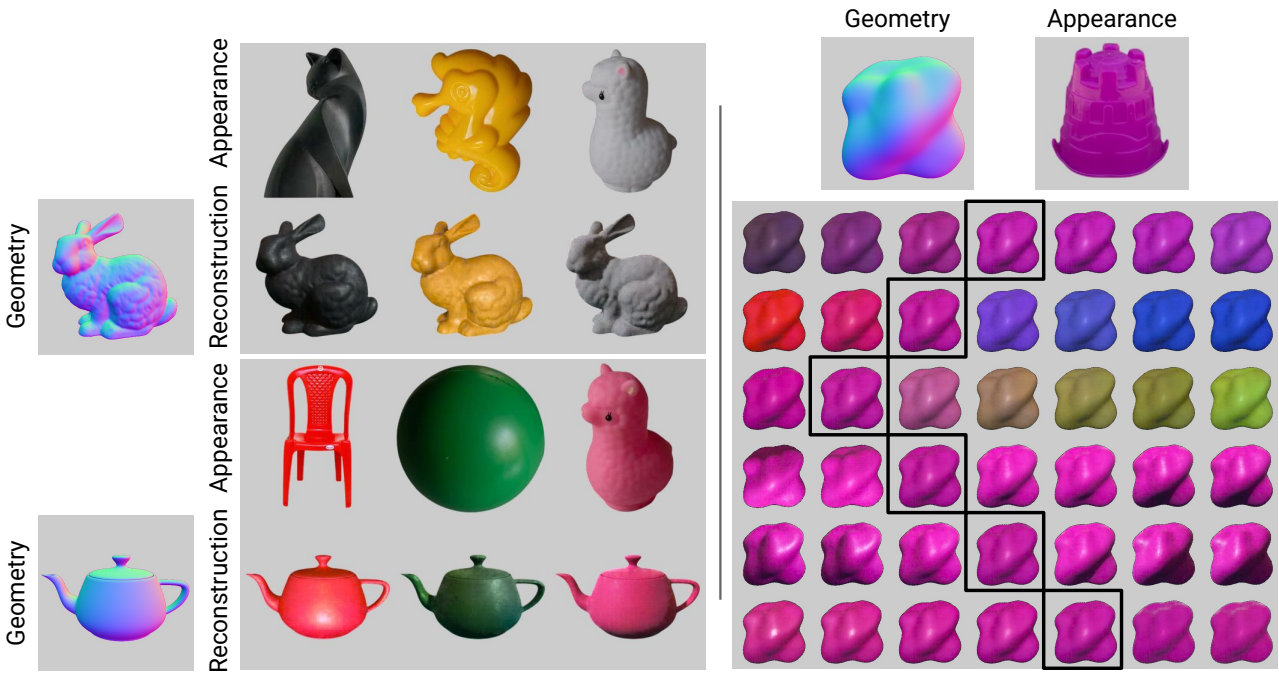

**Figure 4:** ***Left:*** *For six real-world, photographed objects (rows "Appearance", background has been masked), we encode their appearance with our model and decode it with a given normal map ("Geometry"). We see how the reconstructed objects (rows "Reconstruction", showing bunnies and teapots) exhibit the same appearance as their corresponding images, illustrating the ability of our model to encode appearance.* ***Right:*** *Posterior traversals of the latent space for the input image shown, reconstructed with the* blob *normal map (see text for details). Black boxes mark the appearance reconstructed from the input image.*

Finally, we evaluate the ability of our model to encode the material appearance of a given image, *and* to modify such appearance along the dimensions of our latent space. Fig. 4, left, shows, for six real-world, photographed objects (the background has been masked), the result of encoding them with our model and decoding them using the normal map shown. The very similar appearance between the original and the reconstructed result shows the ability of our space to successfully encode the appearance of real objects. In Fig. 4, right, we also encode an input image of a real object into the latent space, obtaining its feature vector, and we then sample this space in each dimension, akin to what was done in the prior traversals plot in Fig. 3, but with the posterior, leading to *posterior traversals* plots. We can see how the material appearance of the input image is correctly captured (black boxes), and how we can modify the original appearance in a controlled manner by traversing the dimensions of the latent space. However, despite the great performance shown by the FactorVAE encoder in successfully distilling a disentangled and interpretable material appearance representation from the input image, the decoder pipeline proves insufficient when generating images of geometries very different from those seen during training (see supplementary material (S3.3) for more details). This highlights the need of a more powerful framework that translates the feature vector obtained into a final image, which is explored further in Sec. 5.

### 4.3. Dimensionality of the Latent Space

Our 6-dimensional space is highly compact, as compared to other alternatives in the literature (e.g., CLIP embeddings [RKH*21] can be 512D (ViT-B/32, ViT-B/16), 768D (ViT-L/14) or 1024D (ViT-H/14), and StyleGAN-based autoencoders [KLA19] are 512D and higher), facilitating interpretability. We explore the effect of modifying the dimensionality of our space, aiming to keep a balance between interpretability and disentanglement: larger latent spaces tend to dilute information in more dimensions, penalizing interpretability, while smaller ones need to embed the same information in a more constrained latent space, which often leads to worse disentanglement.

We analyze models from three to ten dimensions in the latent space, including quantitative metrics in Fig. 5 and qualitative results in Fig. 6, in which we show: (1) the factors captured in each dimension of the latent space, by computing the *prior traversals* plots, and (2) the informativeness of the space, by plotting the dimension-wise KL loss during training. Higher values in this plot represent that the learned distribution is different from the standard normal distribution $N(0,1)$, and thus are storing more information. Our model with six dimensions achieves the best balance between interpretability and disentanglement, as represented by MIR (higher is better) and MIS (lower is better) metrics, respectively (Fig. 5). This six-dimensional space is also more visually disentangled and interpretable, while not including uninformative dimensions (Fig. 6).

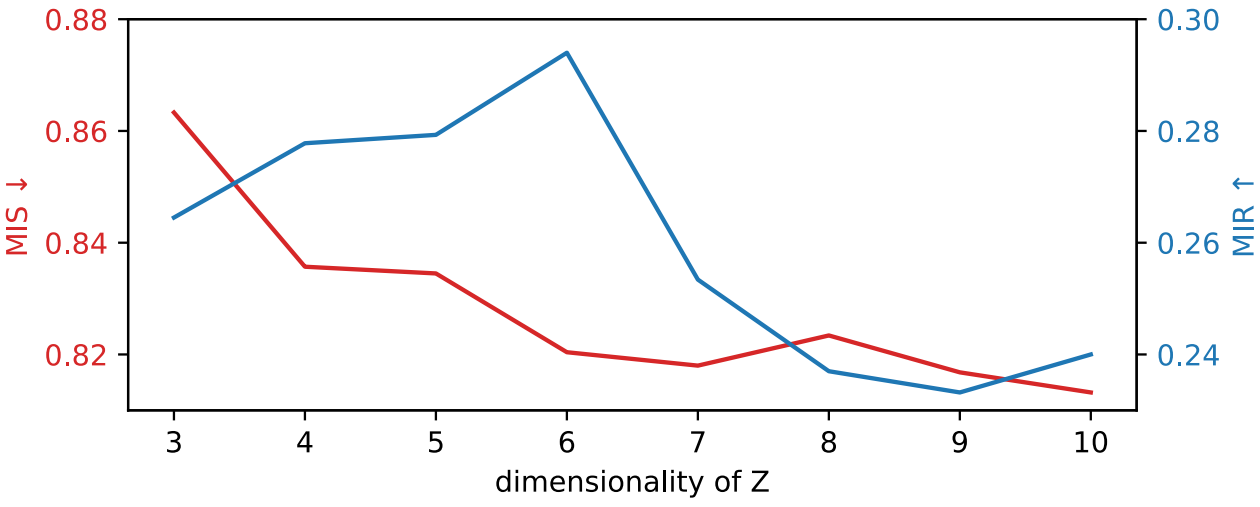

**Figure 5:** ***Latent space dimensionality analysis.*** *Evolution of metrics for interpretability (MIR, higher is better) and disentanglement (MIS, lower is better) for models whose latent space dimensionality ranges between 3 and 10. Our 6-dimensional space achieves the best balance between these two properties.*

**Table 2:** ***Ablation studies.*** *Our final model performs best both in terms of interpretability, as indicated by the MIR score, and reconstruction quality, measured with PSNR (see text for details).*

| | MIR↑ | PSNR↑ |
|---|---|---|
| (a) Summation | 0.2497 | 29.12 |
| (b) Maximum $\beta = 1$ | 0.2306 | 30.08 |
| (c) No $\beta$ annealing | 0.2130 | 30.36 |
| Without Normals | 0.2560 | 28.60 |
| *Ours* | **0.2940** | **30.80** |

### 4.4. Ablation Studies

We evaluate the effectiveness of our design decisions by running ablation studies. We include here ablations on our proposed modifications of the loss function to tackle posterior collapse (Sec. 3.1.1), and on the use of normal maps to enforce the encoding of appearance (Sec. 3.1.2). For additional ablations on the reconstruction loss and normal map resampling, please refer to the supplementary material (S4).

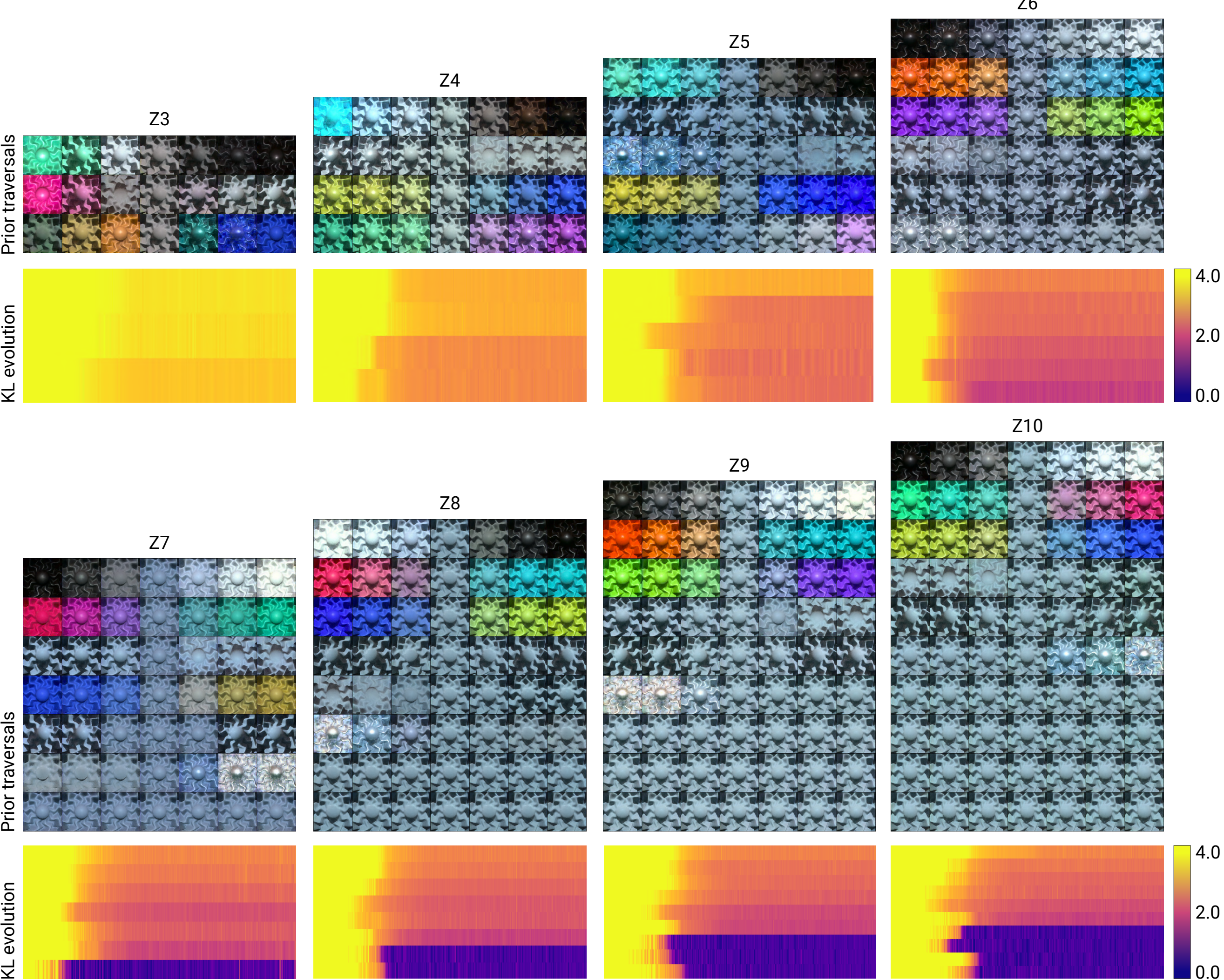

**Figure 6:** *Prior traversals and KL evolution plots of each model trained with latent space dimensionalities ranging from 3 to 10. KL evolution plots shows the dimension-wise evolution, during training, of the KL distance between the learned distributions and the standard normal distribution used to regularize (Sec. 3.1.1). The higher KL distance, the more information is stored in a given dimension.*

#### 4.4.1. Loss Function

We evaluate the effectiveness of the changes we propose with respect to the vanilla FactorVAE loss by training the following alternatives: (a) a model using the default summation instead of our proposed norm (i.e., $n = 1$ in the second term of Eq. 1), (b) a model with $\beta = 1$, and (c) a model without annealing on the $\beta$ parameter with our default $\beta = 2$. Table 2 shows how systematically removing our modifications leads to models with reduced interpretability, as represented by lower MIR values.

#### 4.4.2. Use of Normal Maps

In order to facilitate the disentanglement between appearance and geometry, we guide the reconstruction done by the model's decoder with normal maps. We ablate this modification in Table 2, training a model without this geometry guidance. As expected, we observe how leaving out this information diminishes reconstruction quality of the model, as measured by the PSNR metric. Additionally, the absence of normals leads to reduced interpretability (lower MIR) by requiring geometry to be encoded as an additional factor of variation.

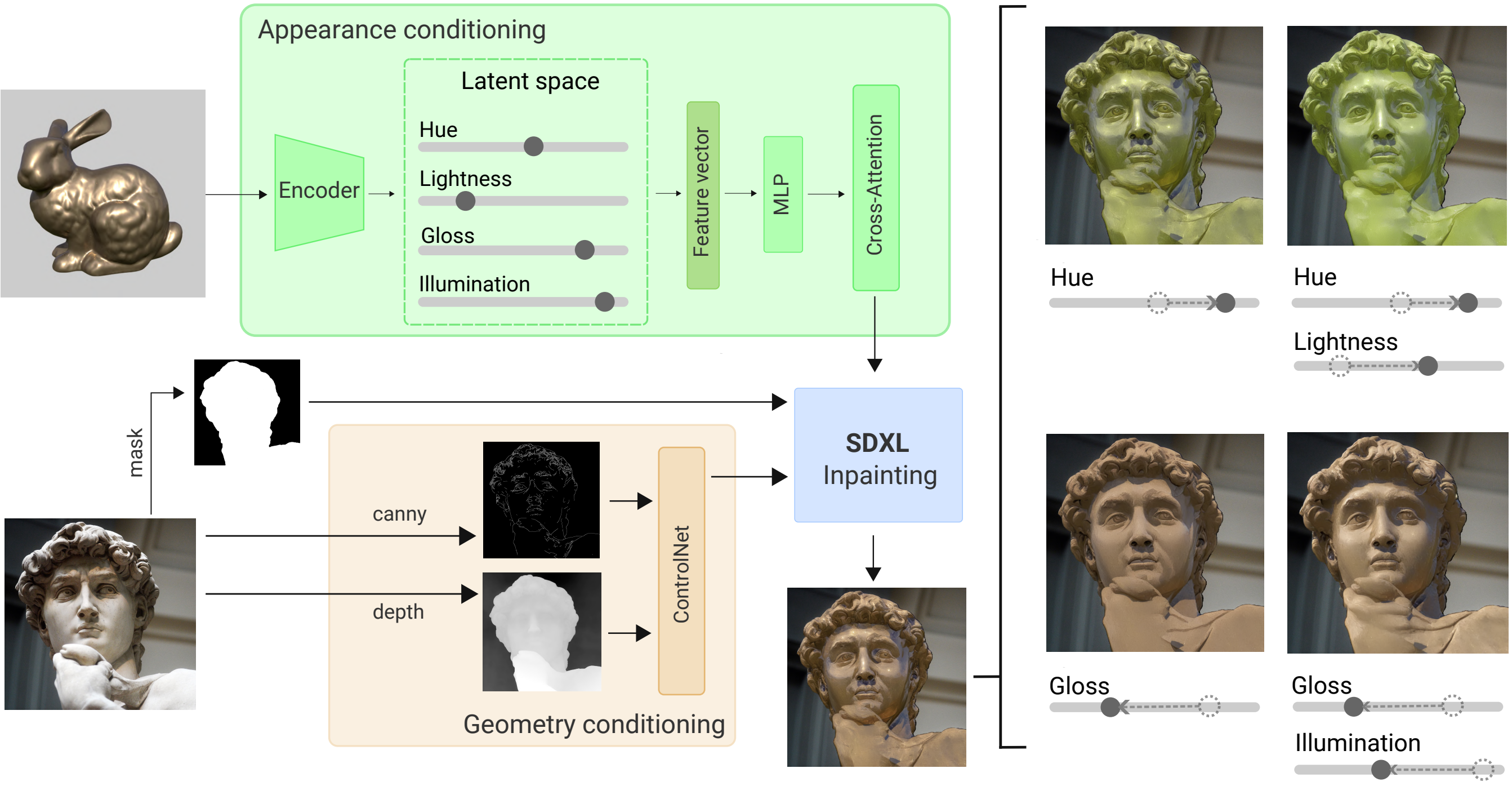

**Figure 7:** ***Diffusion-based pipeline for proof-of-concept applications of our space.*** *Our proposed pipeline uses two branches to condition the diffusion-based generative process with Stable Diffusion XL (SDXL). The appearance conditioning branch leverages our encoder to produce a 6D feature vector representing the desired appearance. This representation can be further edited along each of the six dimensions if desired, providing fine-grained control over the final appearance. The geometry branch leverages ControlNet to condition generation through Canny edges and depth information. We show here appearance transfer from an input image (*bunny*) to a target one (*David*), as well as editing along different dimensions of the latent space (right). Other uses, such as direct editing or selective transfer, are also possible.*

## 5. Applications: Appearance Transfer and Editing

We leverage our compact and controllable latent space (Sec. 3) for two applications: *appearance transfer*, which involves transferring the appearance of one or more reference exemplars to a target one, and *editing*, which modifies the visual appearance of an object in image space. We showcase these proof-of-concept applications by using our representation to condition a diffusion-based pipeline (Sec. 5.1), and evaluate its advantages and limitations (Sec. 5.2).

### 5.1. Diffusion-Based Pipeline

We design a diffusion-based pipeline that uses two sources of information as input: a *geometry reference* image, which defines the target geometry of the object, and an *appearance feature vector* in our latent space, which specifies the desired appearance to be applied to the target geometry. An overview of our pipeline is shown in Fig. 7. We use pre-trained latent diffusion model, RealisticVisionXL4.0, which is a fine-tune of the base model Stable Diffusion XL designed for photo-realism. We add our sources of information to condition the generative process through two distinct branches: an appearance conditioning branch and a geometry conditioning one.

The *appearance conditioning* branch encodes the material and illumination information from one or more reference images into an appearance feature vector by using our encoder (Sec. 3), effectively performing appearance transfer. Alternatively, one can directly sample the latent space to generate such vector. Further, the user can navigate the space, enabling controlled fine-grained adjustments to the generated images (appearance editing). Our appearance encoder, trained as explained in Sec. 3, is kept frozen during the training of the diffusion pipeline. We integrate the information of the appearance feature vector by training an IP-Adapter [YZL*23]. Following the IP-Adapter implementation, we plug an external network via a cross-attention mechanism, and train it by minimizing the original Stable Diffusion loss [RBL*22]. Only the weights of the external network are updated during training, in order to preserve the generative capabilities of the base model.

The *geometry conditioning* branch takes an image as input and incorporates the target geometry information into the diffusion pipeline via a combination of pre-trained ControlNets [ZRA23], designed to process Canny edges and depth maps. Depth information helps in transferring the general structure of the shape, while the Canny edges map allows to preserve the high-frequency details. For inference, we automatically obtain the depth map from the input image using a single-view depth predictor [KOH*24].

Finally, we follow previous work [CSM*24] and use an inpainting base model during inference, which restricts the generation to the area of the object, keeping the background intact. Please refer

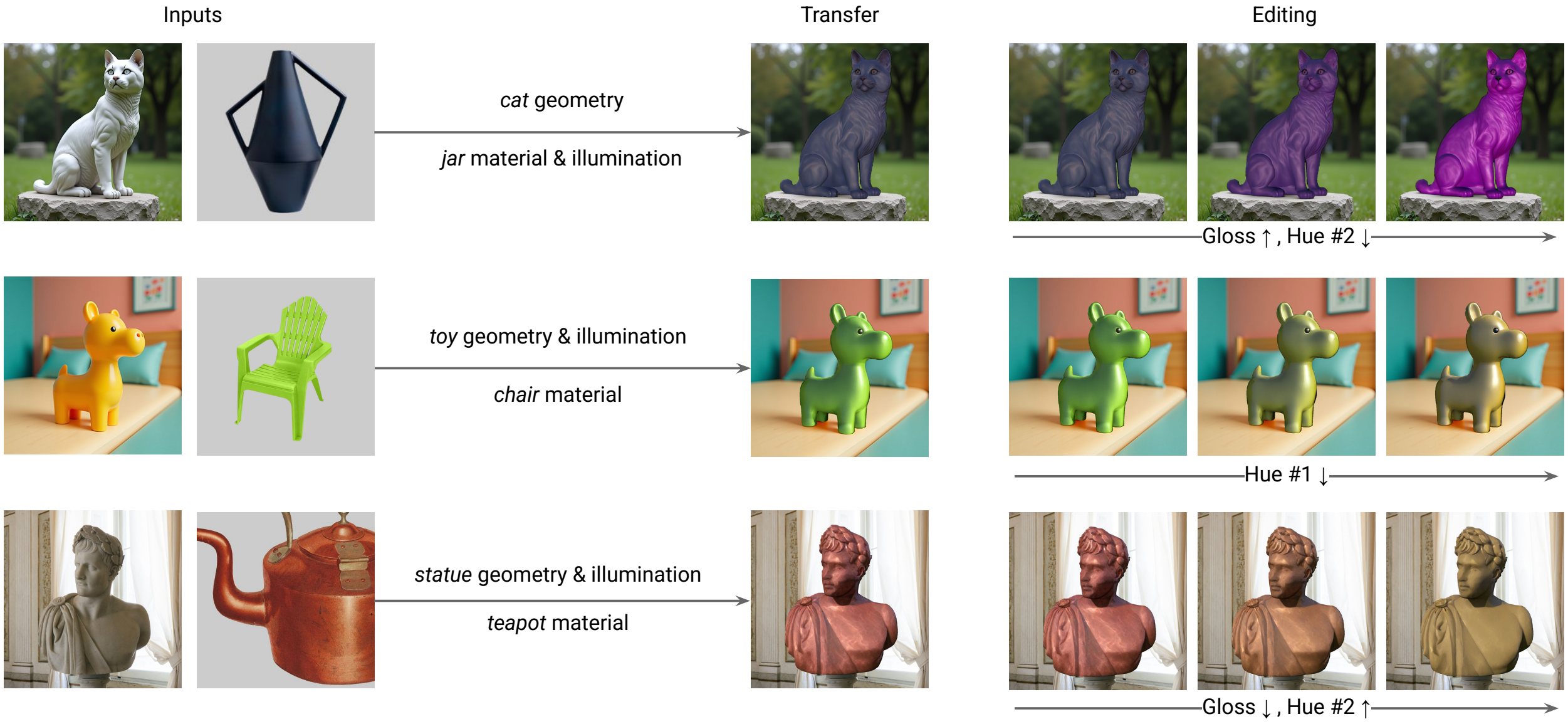

**Figure 8:** ***Appearance transfer and editing results*** *of our pipeline for real-world images leveraging our disentangled appearance representation. In each example, we use two input images, with the leftmost one as the target. We selectively transfer material and/or illumination properties from the other input image (background has been masked out) by encoding them into our latent space. We can further modify the appearance in this latent space, leading to fine-grained editing (right).*

to the supplementary material for full implementation details and ablations of our pipeline (S1.2).

Our pipeline is therefore versatile, and can be used for different applications depending on how the inputs are configured. Fig. 7 illustrates a canonical use case in which the appearance of the *bunny* is encoded with our model and transferred to the image of Michelangelo's *David* with our diffusion pipeline, performing appearance transfer. Once encoded, given the disentanglement and high controllability of our latent space, the user can easily modify the appearance vector and re-generate the image, thus performing fine-grained appearance editing, as shown. Our pipeline is however flexible to other use cases, such as introducing the same image for both appearance and geometry conditioning to perform direct editing (Fig. 11), selective transfer by extracting different factors of appearance from different input images (Fig. 8 and Fig. 1, right), manually defining the appearance of an object by sampling our latent space, or by interpolating in this space (Fig. 9).

### 5.2. Experiments

We evaluate our appearance-aware diffusion pipeline, showcasing applications of our disentangled appearance representation in different use cases. Additional results of our diffusion-based pipeline can be found in the supplementary material (S6).

#### 5.2.1. Qualitative Evaluation

The fact that our space is disentangled enables integrating appearance information from two or more source images when performing appearance transfer. This is particularly useful for the illumination: one can perform material transfer from a source to a target image while keeping the illumination of the target (e.g., Fig. 8, second and third rows). Another example of selective transfer (i.e., transferring different dimensions from several images) beyond illumination dimensions is shown in Fig. 1 (right).

We include further results of our pipeline for both appearance transfer and editing tasks in Fig. 1 (center) and Fig. 8, using images not seen during training. In Fig. 8, for each row, we use two reference images as input (left) and perform appearance transfer between them (middle). Then, we modify this result by traversing the relevant dimension(s) of our latent space, performing disentangled editing (right). We show how our pipeline effectively captures the target appearance achieving realistic results, even when the reflectance properties are very different between the two input images (e.g., the copper glossy material on the statue, third row). The second and third rows further illustrate examples of integrating appearance information from two source images, as we extract the illumination from the feature vector of one of the images (two dimensions) and the material from the other (four dimensions). Despite being trained on synthetic data, our material and illumination appearance transfer generalizes well to real photographs. For editing, Fig. 8 shows how we can perform fine-grained modifications for different attributes while maintaining the identity of the image. We include an example modifying a single attribute (second row) to show the high disentanglement of our latent space. However, editing multiple attributes at once is also natively supported by our pipeline, as shown in the first and third rows.

Our pipeline can also be used to interpolate between two appearance feature vectors, as shown in Fig. 9. Despite significant differences in properties such as hue or gloss between both reference materials, the results exhibit realistic, smooth transitions.

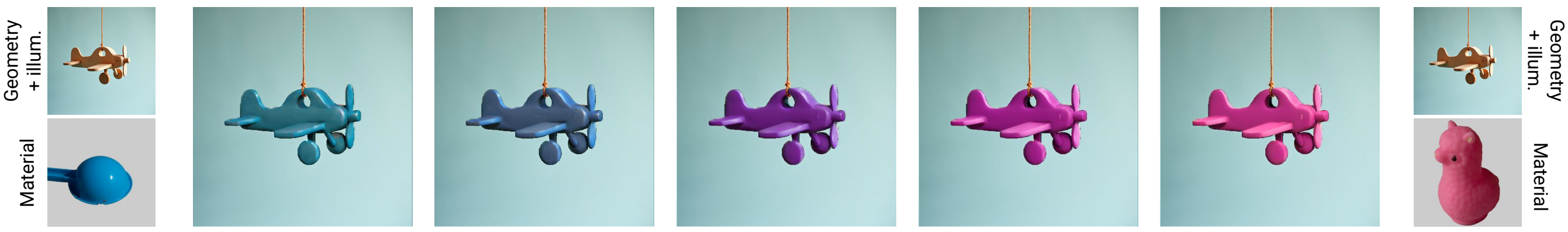

**Figure 9:** ***Interpolation between two materials in our latent space.*** *The two ends of the progression show the result of transferring the material of two real-world objects, a blue glossy spoon and a pink rough llama, to a target geometry. All results follow the illumination of the target image. Progressively traversing our latent space results in an intuitive change of the appearance.*

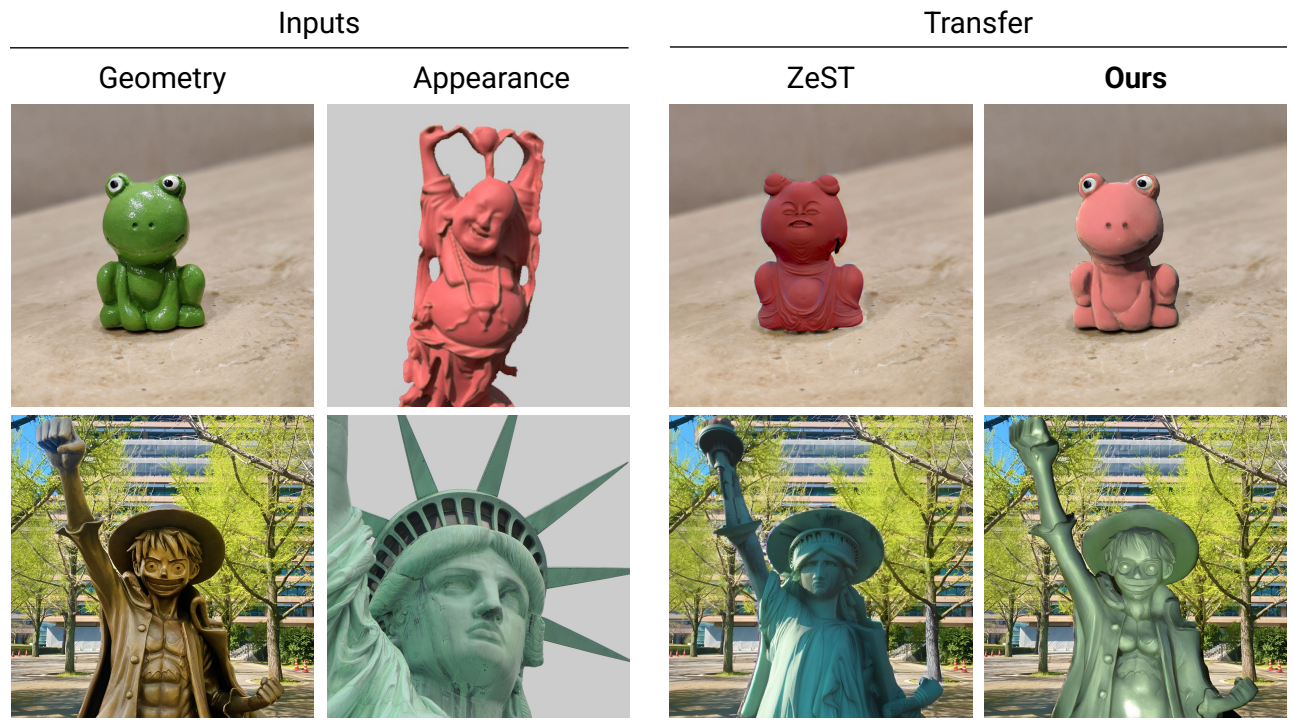

**Figure 10:** *Examples highlighting the disentanglement issue in ZeST [CSM*24] for appearance transfer. The appearance reference image of the first row is a rendering of a homogeneous buddha, while the second row uses a real photograph of the Statue of Liberty. In contrast to our method that properly disentangles appearance information from geometry, the ZeST results include geometric information being transferred to the output image.*

#### 5.2.2. Benefits and Limitations for Transfer

The current state of the art in material transfer is ZeST [CSM*24]. An important difference with our approach lies in their use of a pre-trained *semantic* image encoder which projects the reference appearance image into the space of CLIP [RKH*21]. This encoder is more general than our appearance encoder, enabling ZeST to handle a wider variety of appearance.

However, unlike our method, CLIP has not been specifically trained to disentangle the appearance and geometry. As a result, some geometric information may be transferred through the appearance branch. Fig. 10 illustrates how this limitation of ZeST manifests in the generation of features such as the Buddha face or the Statue of Liberty's facial expression, which originate from the *appearance* image rather than the *geometry* image. As a result, details of the geometry reference image are modified or lost, and material appearance can be influenced by image semantics (e.g., Statue of Liberty). In contrast to CLIP, our appearance encoder explicitly encourages disentanglement between appearance and geometry, leading to a better performance than ZeST in this aspect (additional results in the supplementary material). Besides, our approach can be used to selectively transfer only certain attributes of a given set of inputs (Fig. 1, right).

#### 5.2.3. Benefits and Limitations for Editing

We further evaluate our appearance editing task in Fig. 11, by comparing our results with two state-of-the-art editing methods in image space with publicly available code (Subias and Lagunas [SL23] and InstructPix2Pix [BHE23]), to highlight the benefits and limitations of our space. For every source image and method, we show appearance editing results by traversing two appearance factors, one after the other. The method by Subias and Lagunas is a GAN-based architecture trained supervisedly on specific attributes, so a direct comparison is only possible for the gloss attribute. InstructPix2Pix is a general image editing method conditioned on instructional text prompts, not solely targeted to appearance editing. While we lack such generality, compared to InstructPix2Pix, we can achieve finer-grained edits that only affect the desired material, as well as an increased control over the edit. This highlights the benefits of modeling an explicit disentangled latent space: although InstructPix2Pix has higher generative capacity, our approach is more suitable for *controllable* appearance editing. Finally, color shifts are a common challenge in generative models. While our method is not immune to this issue, its design allows for partial correction by manually adjusting the shifted dimensions to achieve the desired appearance, an ability that is often lacking in related works.

## 6. Discussion and Limitations

We show that we can, in a self-supervised manner and without the need for human-annotated data, learn a disentangled, interpretable and controllable space of appearance. We carefully evaluate the capabilities of such space, as well as alternative design decisions.

To illustrate potential uses of our space, we use it to condition a diffusion-based generative pipeline, enabling proof-of-concept applications: appearance transfer from one or more images, fine-grained editing, and interpolation within the space. We show these for real images, even though our models have been trained on synthetic data. We compare to existing dedicated material transfer or image editing methods for our sample downstream tasks. Despite the higher generative capacity of targeted methods, which we do not aim to surpass, our comparisons highlight the potential benefits of our space (i.e., fine-grained control and interpretability). Moreover, our representation of appearance could have alternative applications, such as generating a certain appearance from scratch by adjusting the different dimensions (exposed, e.g., as sliders), or to be used as a descriptor for attribute-based retrieval in large image databases.

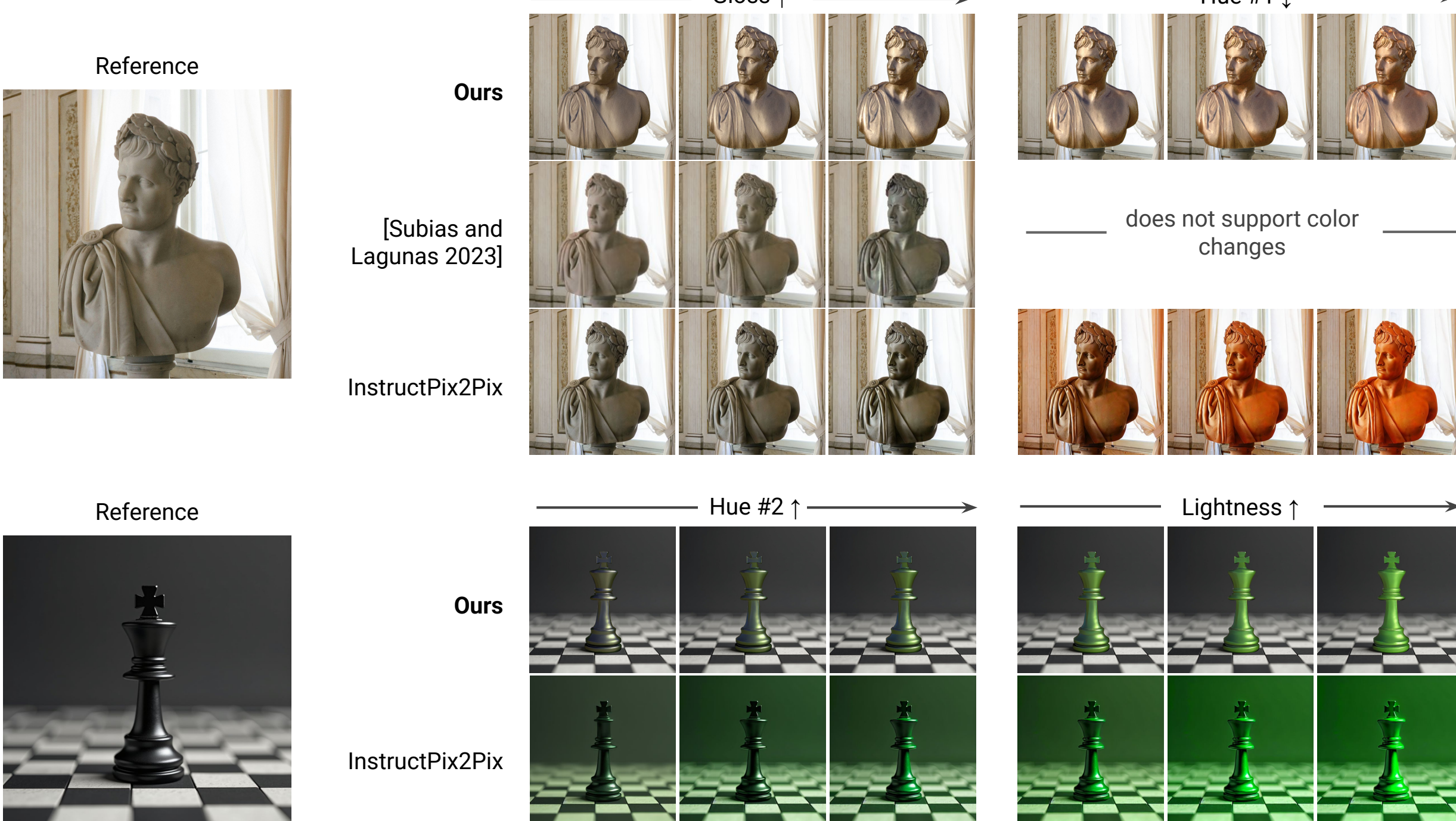

**Figure 11:** *Comparison of the appearance editing feature of our pipeline with two state-of-the-art solutions. Using a reference image (left), we show the progression of sequentially editing two attributes (right). The disentanglement of our latent space and the design of the diffusion pipeline allows for an intuitive and precise editing of the identified attributes.*

Our appearance encoding model is limited to the range of appearances it was trained with, namely homogeneous, opaque materials and illuminations that do not exhibit very high frequency or strongly colored lighting. Fig. 12 shows reconstruction results when trying to encode samples with out-of-distribution, high-frequency illumination. Given the self-supervised nature of the approach, extension to a wider set of appearances increasing the training dataset remains as future work.

A promising direction for future research involves investigating the benefits of increasing the level of supervision in the training process. The current approach employs soft disentanglement constraints via the total correlation (TC) term in the loss function (Eq. 1). Still, we evaluate the disentanglement and interpretability of our space (Tab. 1), obtaining successful results. In future work, imposing hard constraints directly within the network architecture [KWKT15] could enable the enforcement of higher disentanglement in the learned latent dimensions. This could be implemented by fixing our extrinsic factors (geometry and illumination) and learning the intrinsic properties of the material.

Further, while the dimensions of our space are interpretable and controllable, they are not necessarily perceptually-linear, since no supervision enforces this. Although they are independent factors determining appearance, they may not be expressive enough and artists could prefer to control more, or alternative, dimensions.

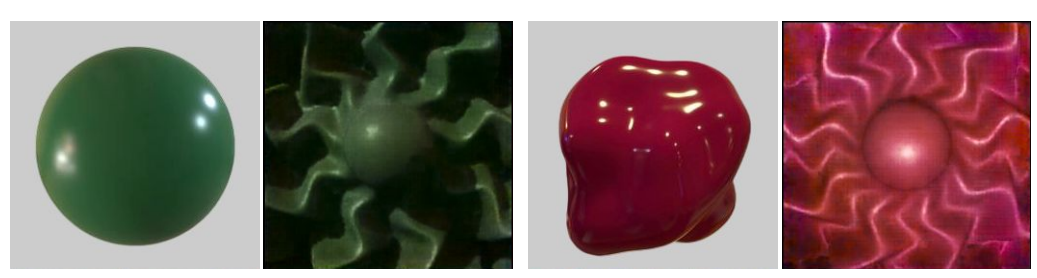

**Figure 12:** ***Limitations.*** *Examples of the behavior of our autoencoder when reconstructing out-of-distribution appearances. The* sphere *and* blob *samples are illuminated with high-frequency lighting. Reconstructing the* Havran *geometry using their respective embeddings, the model struggles and cannot recover appearance.*

Alternatively to our work, using supervision could allow to control specific factors of variation (e.g., varying parameters of an analytical BRDF model). However, it would necessarily require factors defining appearance to be defined *a priori*, which would introduce some bias, and could guide the space to learn the notion of, e.g., specular from the underlying model used. Instead, we have focused on self-supervised learning to investigate whether a meaningful appearance space can emerge from diverse realistic images, whose appearance factors of variation are a priori *unknown*. We hope our work inspires further exploration of self-supervised learning approaches to uncover the underlying factors that shape our perception of appearance.

**Acknowledgments**

This work has been partially supported by grant PID2022-141539NBI00, funded by MICIU/AEI/10.13039/501100011033 and by ERDF, EU, and by the European Union's Horizon 2020 research and innovation programme under the Marie Skłodowska-Curie grant agreement No. 956585 (PRIME). Julia Guerrero-Viu was also partially supported by the FPU20/02340 predoctoral grant. We thank Daniel Martin for his help designing the final figures, Daniel Subias for proofreading the manuscript, and the members of the Graphics and Imaging Lab for insightful discussions. We also thank the people who participated in the user study, and the I3A (Aragon Institute of Engineering Research) for the use of its HPC cluster HERMES.

# Supplemental Material: A Controllable Appearance Representation for Flexible Transfer and Editing

Santiago Jimenez-Navarro , Julia Guerrero-Viu & Belen Masia
Universidad de Zaragoza, I3A, Spain

The supplemental material of this paper contains additional results and details not included in the main manuscript for conciseness. It is distributed as follows:

- (S1) Implementation Details of the Trained Models
- (S2) Training Dataset: Additional Details
- (S3) Additional Results of our Appearance Encoder
- (S4) Additional Ablation Studies of our Appearance Encoder
- (S5) Additional Details of our Diffusion Pipeline Evaluation
- (S6) Additional Results and Ablations of our Diffusion Pipeline

## S1. Implementation Details of the Trained Models

In this section we provide further details on the configurations used to train the models discussed in the main text.

### S1.1. Appearance Encoder

The network implementation builds on the work by Dubois et al. [DKLM19], which contains the definition of different VAE-based models. The system is developed using the Pytorch library [PGM*19], and trained on a NVIDIA GeForce RTX 3090. We use the Adam optimizer to optimize both the autoencoder and the discriminator in charge of computing the TC term, with learning rates of $1e^{-3}$ and $7e^{-5}$, respectively. We use a batch size of 150, and exclude the normal map information from the first two deconvolution layers. The model is trained during 2,400 epochs, with a total training time of 106 hours. The $\gamma$ parameter is fixed to a constant value of 6, and we perform an annealing of the $\beta$ parameter, which grows linearly from 0 to 2 during the first 1,000 epochs. For the regularization term, we use n=3.

### S1.2. Diffusion-based Pipeline

We use *RealisticVisionXL4.0* (https://huggingface.co/SG161222/RealVisXL_V4.0) as the base model for our diffusion pipeline. It is a fine-tuning of Stable Diffusion XL, especially aimed at photorealism. Following community recommendations, we train the default architecture of IP-Adapter [YZL*23] for reconstruction on our training dataset. To ensure accurate appearance embeddings during training, we use the same inputs used to train our appearance encoder, at 512x512 resolution. Using object-centered images with the background masked out, we encourage the learned weights of the adapter to store information aligned with the FactorVAE's latent space. Additionally, we effectively remove the influence of the text embeddings by using uninformative prompts, such as *"image"*. Training is done in a NVIDIA A100 GPU with $10^{-4}$ learning rate with batch size of 10. During inference, we use the Diffusers [vPPL*22] implementation of the ControlNet inpainting pipeline for Stable Diffusion XL.

## S2. Training Dataset: Additional Details

We carefully create a training dataset suitable for the disentanglement of appearance in image space. It consists of 98,550 synthetic images rendered with Mitsuba [Jak10]. These images are the result of the combination of the following factors:

- 30 geometries. We have selected geometries of different levels of complexity and realism. The 3D models come from publicly available sources: the Morgan McGuire's Computer Graphics Archive [McG17], Pixar's Renderman (https://renderman.pixar.com/community_resources), 3D Assets One (https://3dassets.one), TurboSquid (https://www.turbosquid.com/), and Polyhaven (https://polyhaven.com/models).
- 365 measured BRDFs obtained from real materials. Among these materials, 266 come from the MERL dataset [MPBM03] (173 of which are edits from the original ones [SCW*21]), 36 come from the RGL dataset [DJ18], and the remaining 63 come from the UTIA dataset [FV14]. During the selection of materials, we aimed for a dataset balanced in terms of appearance attributes.
- 9 lighting conditions. We create the different illuminations by systematically rotating the Green Point Park environment map (https://polyhaven.com/a/green_point_park) at angles -50º, -25º, 0º, 25º, and 50º in the X-axis, and 15º, 0º, -15º, -50º, and -70º in the Y-axis.

Fig. 20 shows a representative set of the aforementioned training dataset: the first three rows contain samples of the 30 geometries used, rendered with the same material and illumination. Rows four, five, and six include samples featuring the *bunny* geometry, rendered with a representative set of the materials used. Finally, the last row contains the nine illuminations.

## S3. Additional Results of our Appearance Encoder

This section contains additional results of our model designed to encode appearance in image space. For details on how this model is built, please refer to the main text.

### S3.1. Traversability and Continuity of the Latent Space

In the main text (Fig. 3), we include the *prior traversals* plot of our 6D latent space, highlighting its disentanglement and interpretability properties. Further, and thanks to these properties, we study the result of combining, two by two, these disentangled dimensions. Fig. 13 contains the result of combining three different pairs of dimensions. Thanks to the inherent continuity of VAE-based meth-ods, we see how combining the identified dimensions results in progressive changes in the appearance of the *Havran* geometry.

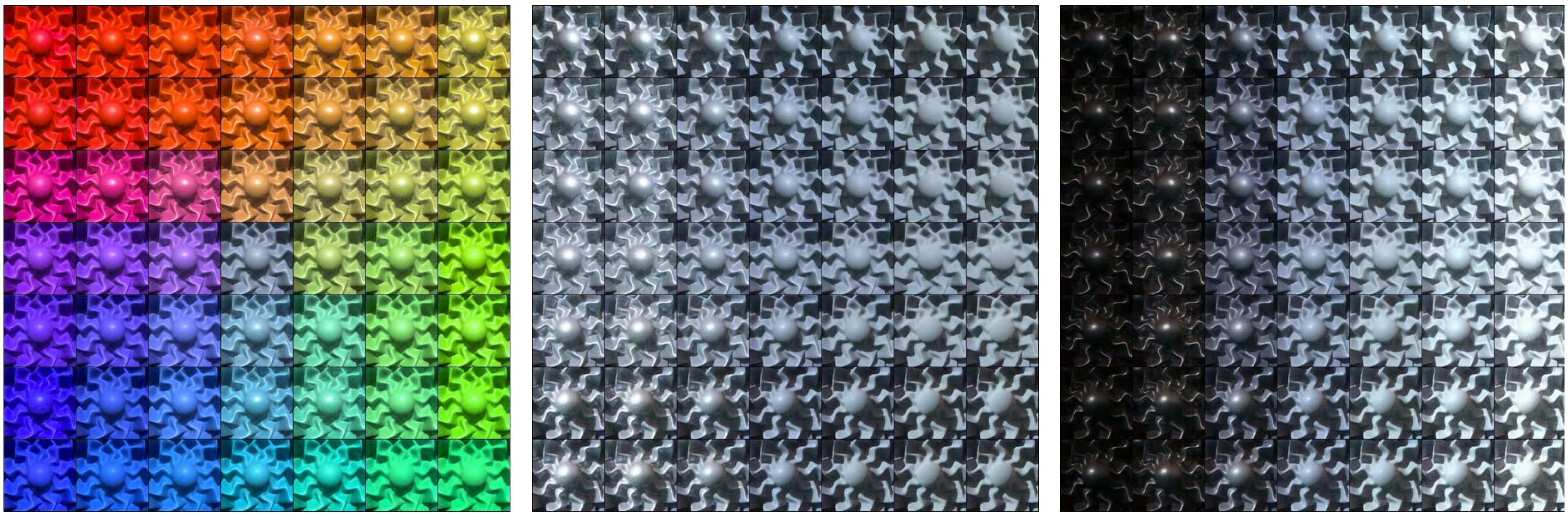

**Figure 13:** ***Grid plots****, showing the appearance reconstructed by the decoder when lineally combining pairs of dimensions of the latent space. We show: (left) hue #1 and hue #2, (center) light dir. #2 and gloss, and (right) light dir. #2 and lightness. Interestingly, the model has automatically learned to represent the hue in two dimensions that are perpendicular in the chromatic circle (left), without any explicit supervision.*

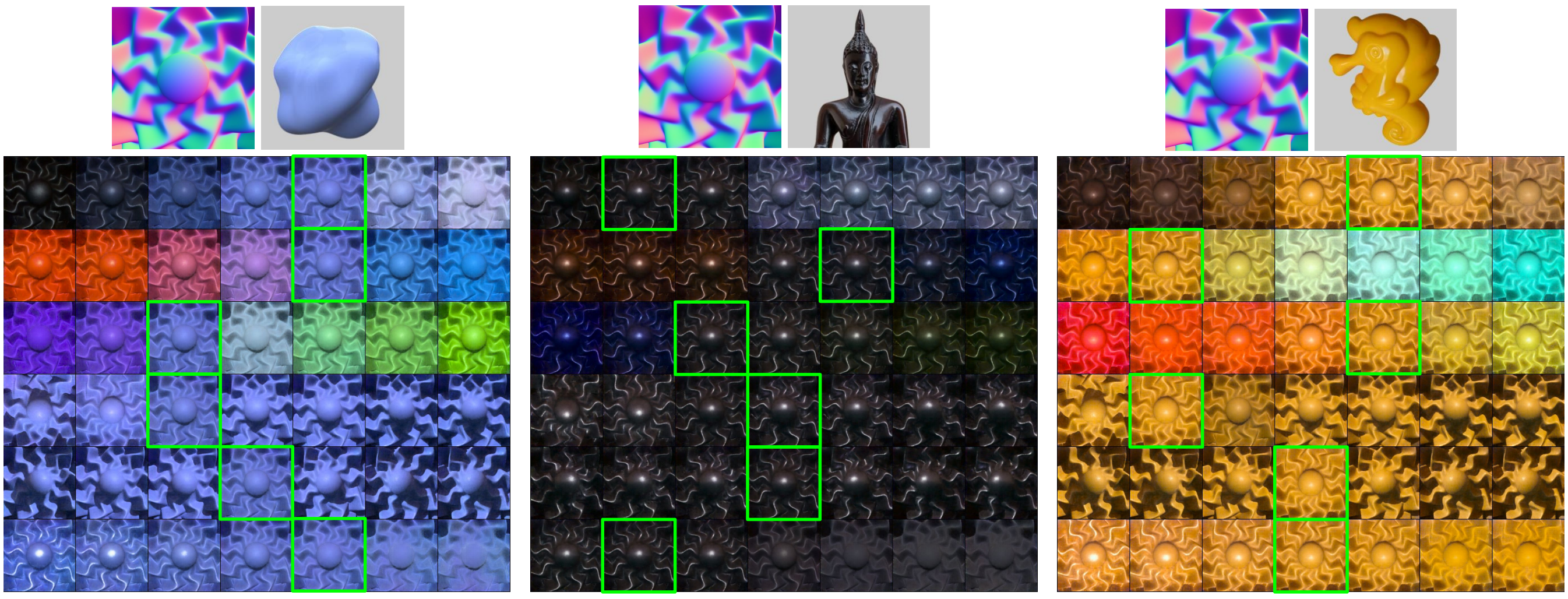

**Figure 14:** *Additional results of the* ***posterior traversal*** *plot. We use as reference three unseen samples, reconstructing the Havran geometry.*

### S3.2. Additional Posterior Traversals

We include in Fig. 14 additional results of the *posterior traversals* plot (see main text, Fig. 4, right). Here, we compute posteriors with three additional test samples: a rendered blue rough blob, and two real photographs of a glossy black statue, and a yellow plastic sea-horse, respectively. We use the *Havran* geometry as guidance in the decoder, and highlight in green the samples corresponding to reconstructions of the reference material, akin to what is done in the main text.

### S3.3. Reconstruction of Unknown Geometries

As discussed in the main text (Sec. 4.2), our VAE-based appearance encoder suffers from a limited capacity to reconstruct geometries very different from those seen during training. This generalizability issue is not an unexpected behavior, but it hinders the applicability of such encoder *alone* to reconstruct appearance in unknown scenarios. In Fig. 15 we show the result of reconstructing the real images used in the main text (Fig. 4) with different objective geometries: as reference, we use the *blob* geometry (row two), which is in the training dataset, and then test the reconstructions for three unknown geometries, namely the *monster*, the *cat*, and the *chair* in rows three, four, and five, respectively. Analyzing the results, we see how the model is able to apply the reference appearance to some extent, but struggles in the estimation of illumination cues (i.e. shadows) and high frequency details of the geometries, while generating unexpected artifacts. This issue arises from the use of unknown geometries, which the decoder is not able to properly interpret.

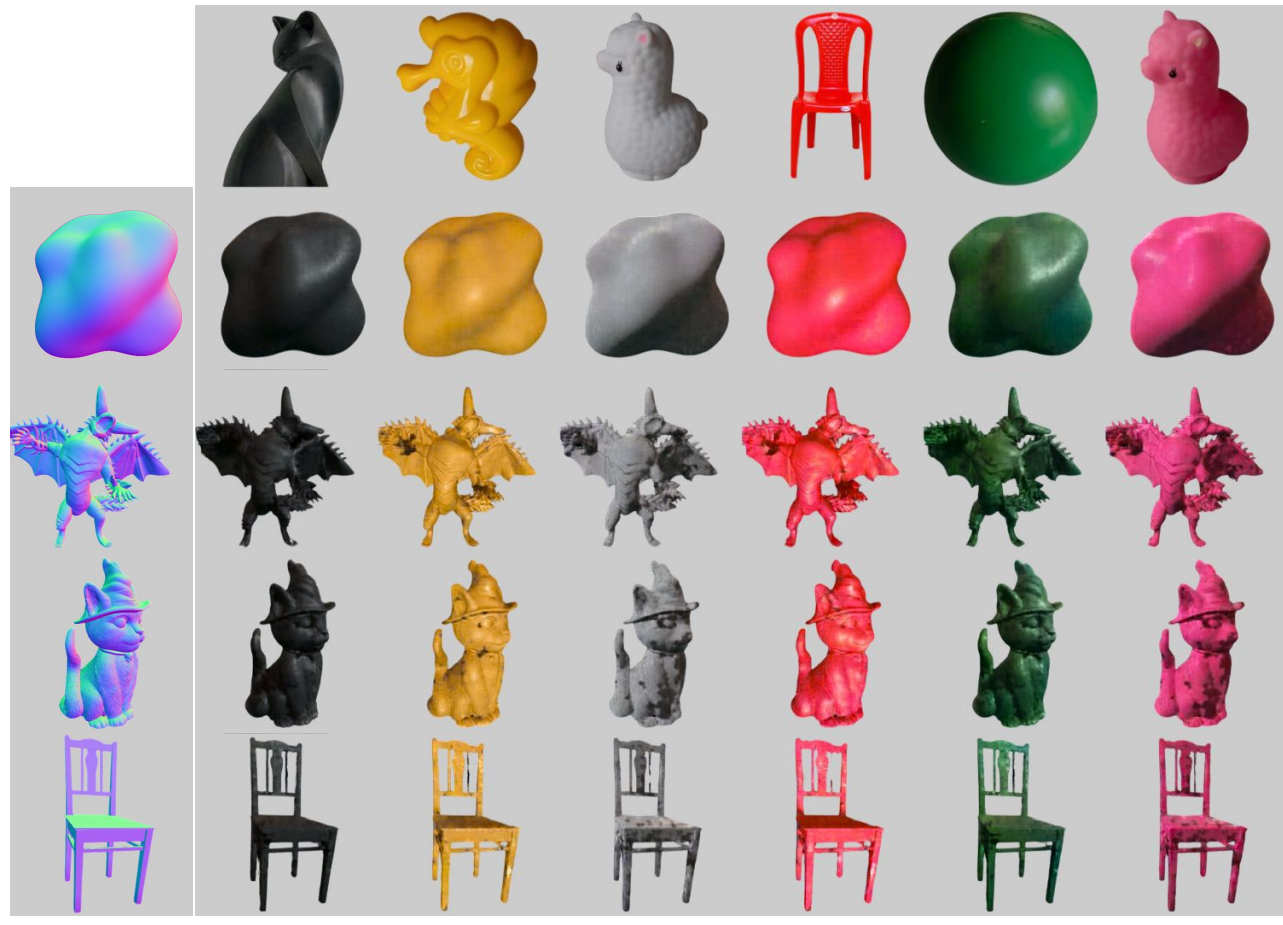

**Figure 15:** *Examples of reconstructions made with different goal geometries. First row depicts the six in-the-wild images used as appearance input. Second row acts as reference, performing a reconstruction using the* blob *geometry seen during training. Rows three, four, and five contain the reconstructions using the test geometries* monster, cat, *and* chair *as reference geometry.*

### S3.4. Robustness of Encoding

To evaluate the robustness of the appearance encoder's embeddings with respect to geometric variation, we conducted an experiment using two distinct test geometries: a *cylinder*, representing a smooth surface, and a *statuette*, characterized by high-frequency geometric detail. All samples were rendered under the *AutoService* environment map. We quantified embedding similarity using the mean cosine similarity (range: [-1, 1]) computed between pairs of embeddings. We observe that the mean similarity between samples of the same material is 0.879, whereas it decreases to 0.321 for samples with different materials. These findings suggest that the encoder consistently maps samples with identical materials to similar regions of the latent space, and vice versa.

## S4. Additional Ablation Studies of our Appearance Encoder

In this section we include additional ablation studies of our appearance encoder, highlighting the thoughtful design of our final model (see main text, Sec. 3.4 for the main ablations).

### S4.1. Reconstruction Loss

During the design of the final model, we tested four commonly used reconstruction losses in autoencoder architectures, namely BCE, Bernoulli, Huber, and L1-smooth [PGM*19]. In the final implementation, we chose to use the L1-smooth function, due to its superior performance in our experiments, as shown in its improved interpretability, measured by MIR, in Table 3). Additionally, the first two reconstruction functions operate at a considerably higher magnitude than the latter two alternatives (Fig. 16), which caused training instabilities.

**Table 3:** *Ablation study on alternative reconstruction loss functions. We include results of interpretability (MIR metric) for four different reconstruction losses.*

| | BCE | Bernoulli | Huber | L1-smooth |
|---|---|---|---|---|
| MIR↑ | 0.1770 | 0.2440 | 0.2622 | **0.2940** |

**Figure 16:** *Ablations. Evolution of different alternative reconstruction losses during training. Note that y-axis is log-scale.*

### S4.2. Normal Map Downsampling Method

In the Sec. 3.1 of the main text, we explain that the geometry guidance in the decoder of the FactorVAE-based architecture is implemented by adequately concatenating the channels to layers of the decoder pipeline. To adequately match the shape of the maps to that of the feature maps of the layers, we need to resize the normal map image. Here we ablate the algorithm used to perform such resizing of the normal map, comparing the nearest neighbors (NN) and bilinear methods. Fig. 17 shows the performance of two models, each one trained with one of the alternatives, when reconstructing a test geometry. We can clearly see how the model that uses the bilinear interpolation achieves a better understanding of the unseen geometry, which is reflected in a reduced (but not inexistent) number of artifacts. This is probably due to a *smoother* reduction of the resolution by the bilinear algorithm, in contrast to the nearest neighbors one.

## S5. Additional Details of our Diffusion-based Pipeline Evaluation

Here we provide details regarding the evaluations performed in the Sec. 5.2 of the main paper. The different real images we used as input of the pipeline come from public datasets [SCW*21; DLC*22], have been AI-generated [Lab23] or come from the following sources:

- Copper teapot (Fig. 1, main, Fig. 6, supp): rawpixel.com.
- Napoleon (Figs. 1, 8, 11, main): Daniel Robert, unsplash.com.
- Green chair 1 (Fig. 1, main): alicdn.com.
- Red chair (Fig. 4, main): imimg.com.
- David (Fig. 7, main): Roberto Serra, news.artnet.com.
- Green chair 2 (Fig. 8, main, Fig. 6, supp): grangecoop.com.

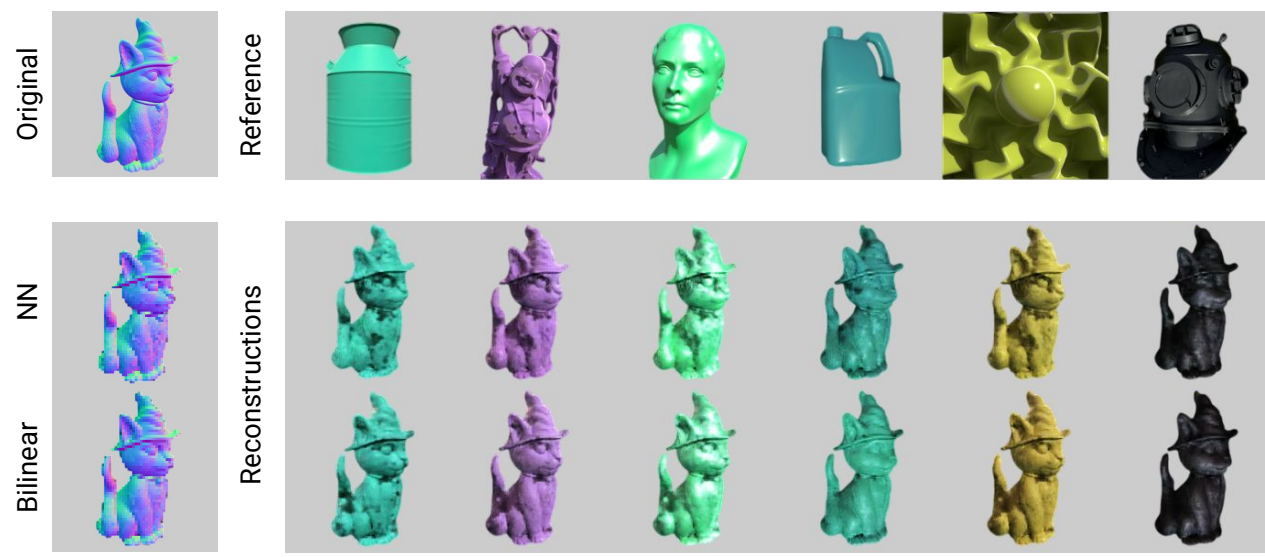

**Figure 17:** *Ablation study on the two alternatives for downsampling the normal map information, namely nearest neighbors (NN), and bilinear. Left: the reference geometry (*cat*), with its respective resulting appearance after applying each of the two algorithms. Right: samples used as appearance reference (first row), with their respective reconstructions using each algorithm (second and third row).*

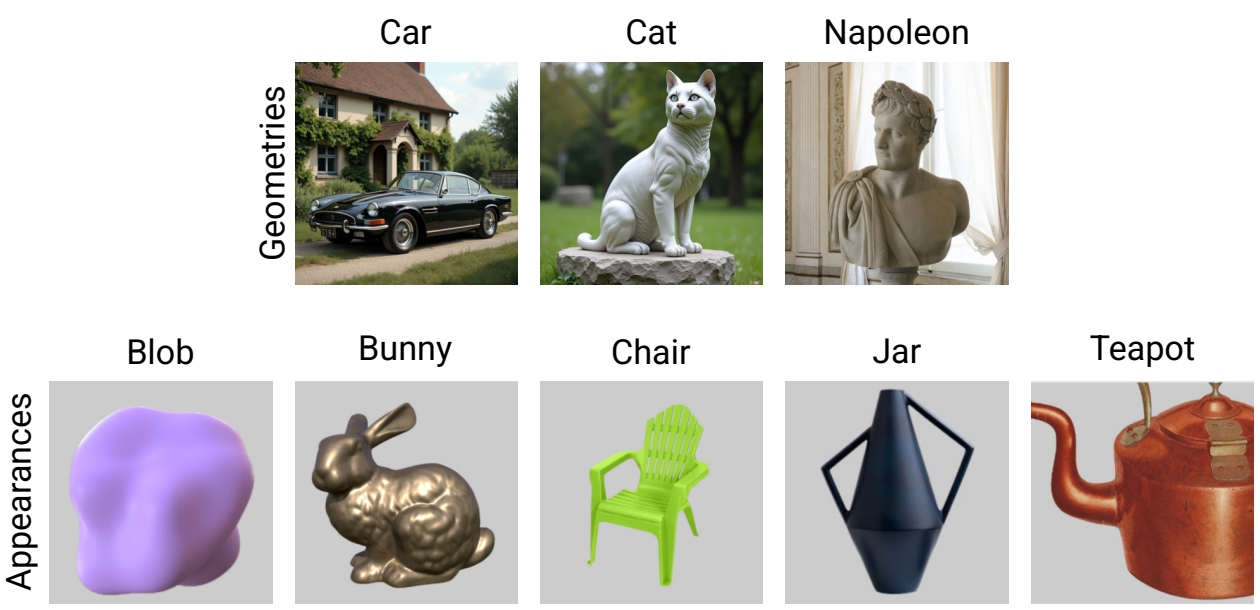

**Figure 18:** *Inputs used for the diffusion-based posterior traversals presented in Figs. 21- 35.*

- Luffy statue (Fig. 10, main): kumamoto.guru.
- Statue of Liberty (Fig. 10, main): Riis2602 Wikimedia.

In the Fig. 8 of the main text, we compare our results with two existing baselines: InstructPix2Pix[BHE23], and Subias and Lagunas [SL23]. For emulating the traversal of our latent space in the InstructPix2Pix pipeline, we used the following configurations:

- For increasing the gloss of the statue, we used the prompt *make the statue shiny*, with a text CFG of 5.0, 6.0, and 7.5.
- For decreasing the hue #2 of the resulting image of the previous stage, we used the prompt *make the statue orange*, with a text CFG of 2.0, 3.0, and 4.0.
- For increasing the hue #2 of the chess piece, we used the prompt *make the chess piece green*, with a text CFG of 0.5, 2.0, and 5.0.
- For increasing the lightness of the resulting image, we used the prompt *make the chess piece light green*, with a text CFG of 1.0, 5.0, and 8.0.

In the case of the Subias and Lagunas' pipeline, we directly used it for the attributes of the comparative handled by their model.

## S6. Additional Results and Ablations of our Diffusion Pipeline

In this section we include additional results obtained with our diffusion-based pipeline, discussed in the Sec. 5 of the main manuscript, and run ablations on some of the design decisions taken in the development of such pipeline.

### S6.1. Posterior Traversals with our Diffusion-based Pipeline

Once trained, our diffusion pipeline is expected to resemble the good properties of our appearance encoder (e.g., disentanglement, interpretability, continuity). To evaluate this, we create a version of the *posterior traversals* plot (main text, Fig. 4, right) with the designed diffusion-based pipeline, using the four dimensions that contain appearance-related features. We use the remaining two dimensions encoding illumination features exclusively to guide the pipeline on the desired lighting conditions, thus we do not perform a traversal on such dimensions. For it, we start from the setup displayed in the teaser (main text, Fig. 1, left), where we apply a rough purple material to the painting of a car. Starting from this appearance transfer (figures marked in green), we systematically vary each of the four dimensions, obtaining as a result the visualization of Fig. 21. Rows follow the same order as the traversals in Fig. 14, namely lightness, hue #1, hue #2, and gloss. Note the overall smoothness of transitions, achieved thanks to the guidance of our custom appearance encoder. When traversing the same dimension as the appearance transfer (e.g. trying to make the paint *even* more purple), the representation may overflow the ranges seen during training, creating saturated results. To further evaluate the robustness of our method, we generate diffusion-based posterior traversals using 15 combinations of the three geometries and five appearances illustrated in Fig. 18. The resulting traversals are presented in Figs. 21- 35. These results encompass a diverse range of appearances and reference geometries, showcasing our method's applicability to real-world scenarios, as well as its limitations in some challenging examples.

### S6.2. Appearance Transfer Evaluation

To further assess the improved disentanglement between geometry and appearance achieved by our custom appearance encoder, we conducted a user study involving 20 participants. Each participant was presented with results from an appearance transfer task and asked to select the output that best fulfilled the objective: "Generate an image with the given geometry and the given material appearance". As a baseline, we used the outputs of ZeST [CSM*24], following the configurations shown in Figs. 21–35. Results showed that our proposed method was preferred in 61.5% of the selections, while ZeST was chosen in 38.5% of the cases.

### S6.3. Ablation Study on the Influence of ControlNets

As described in the main paper, our inference pipeline leverages a combination of two ControlNets [ZRA23] to provide geometry guidance. Each ControlNet must be appropriately weighted, as these weights directly influence the final image. In our work, we use fixed weights of 0.2 for the Depth ControlNet and 0.9 for the Canny ControlNet. To analyze the impact of these weights, we run an ablation study, presented in Fig. 19. As observed, the Canny map may contain some ambiguous edges that require complex in-terpretation during inference. When relying solely on this information (D=0.0, C=1.0), the pipeline may misinterpret certain edges as being outside of the main geometry, leading to artifacts such as the black blob (first row, second column). Conversely, the depth map helps to better identify the geometry but lacks some details of the scene. If only conditioning the diffusion pipeline with the estimated depth map (Fig. 19, last row), high-frequency details are significantly diminished, which is especially noticeable in the sphere (last row, third column). Among the tested configurations, our chosen weights offer the best balance between preserving geometric structure and maintaining high-frequency fine details.

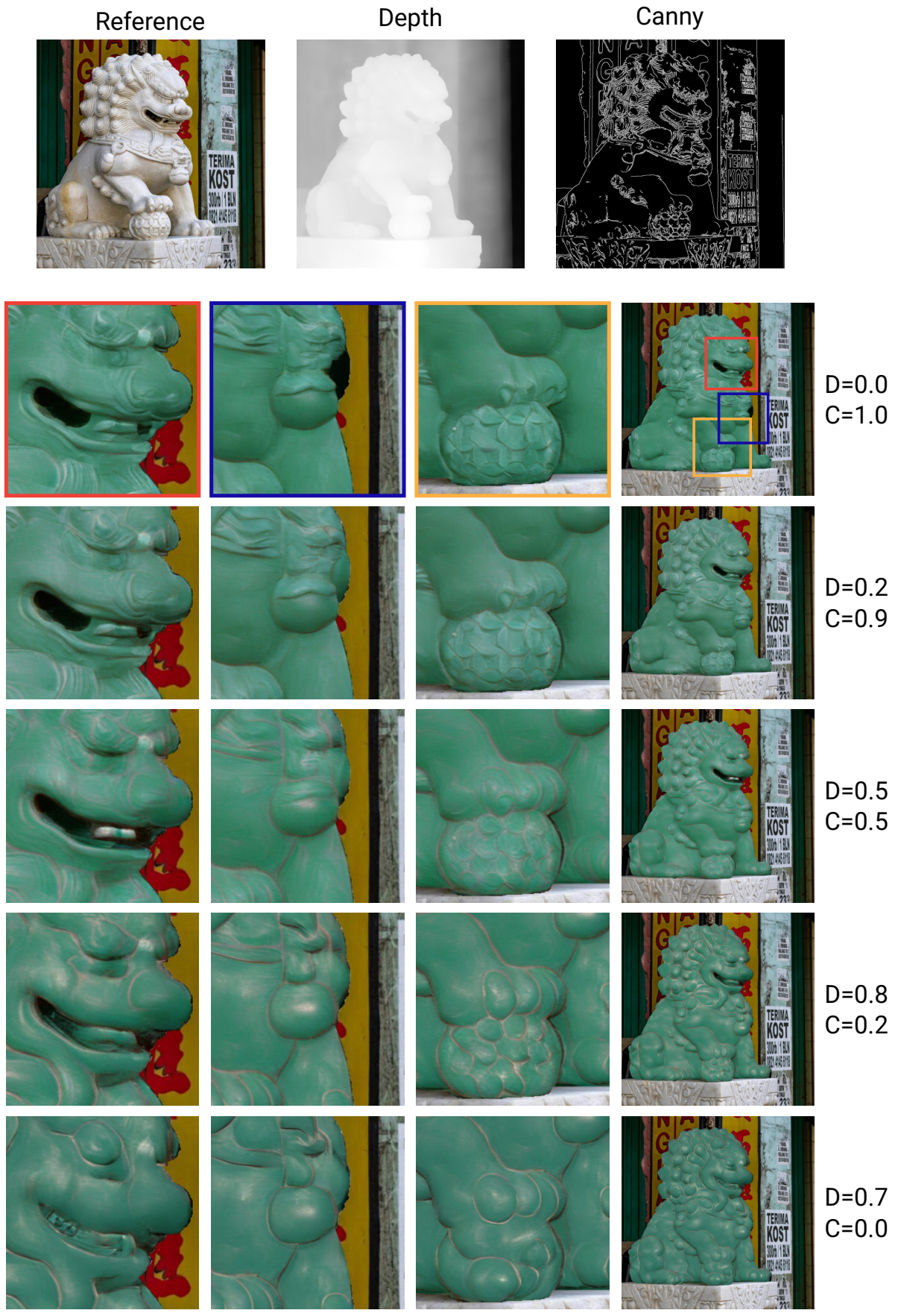

**Figure 19:** *Ablation on the influence of the ControlNet weights over the final result. On top are represented the geometry used as reference, with its respective estimated depth and Canny maps. Each row contains the result of running the inference pipeline with the same custom reference appearance and different weights. We include three close-up zooms for visualization purposes, and the respective weights for the Depth (D) and Canny (C) ControlNets.*

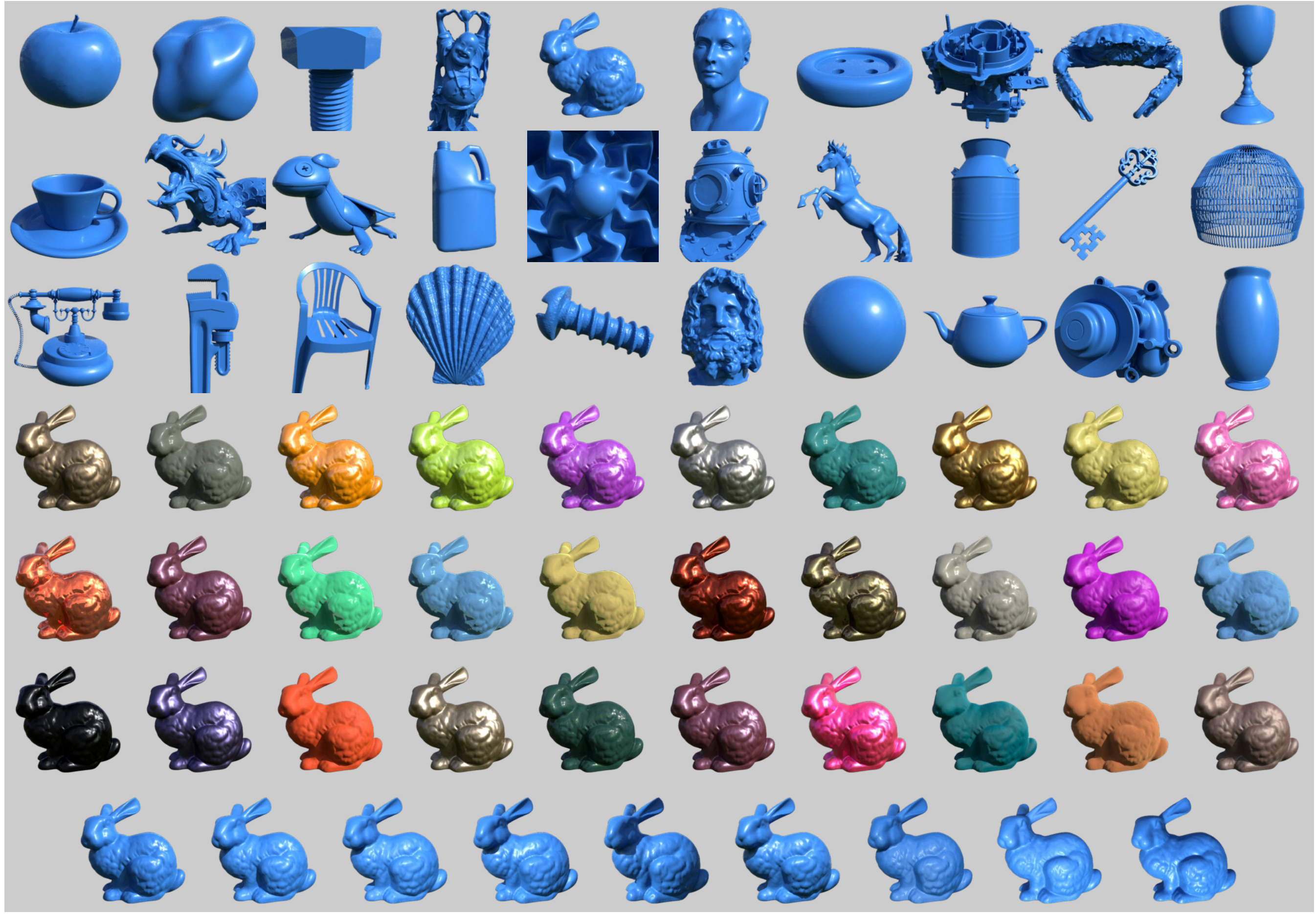

**Figure 20:** *Representative samples of the custom training dataset. Rows one, two, and three contain samples of the 30 geometries used. Rows four, five, and six contain a representative set of the 365 measured materials. Row seven shows the nine illuminations used in the dataset.*

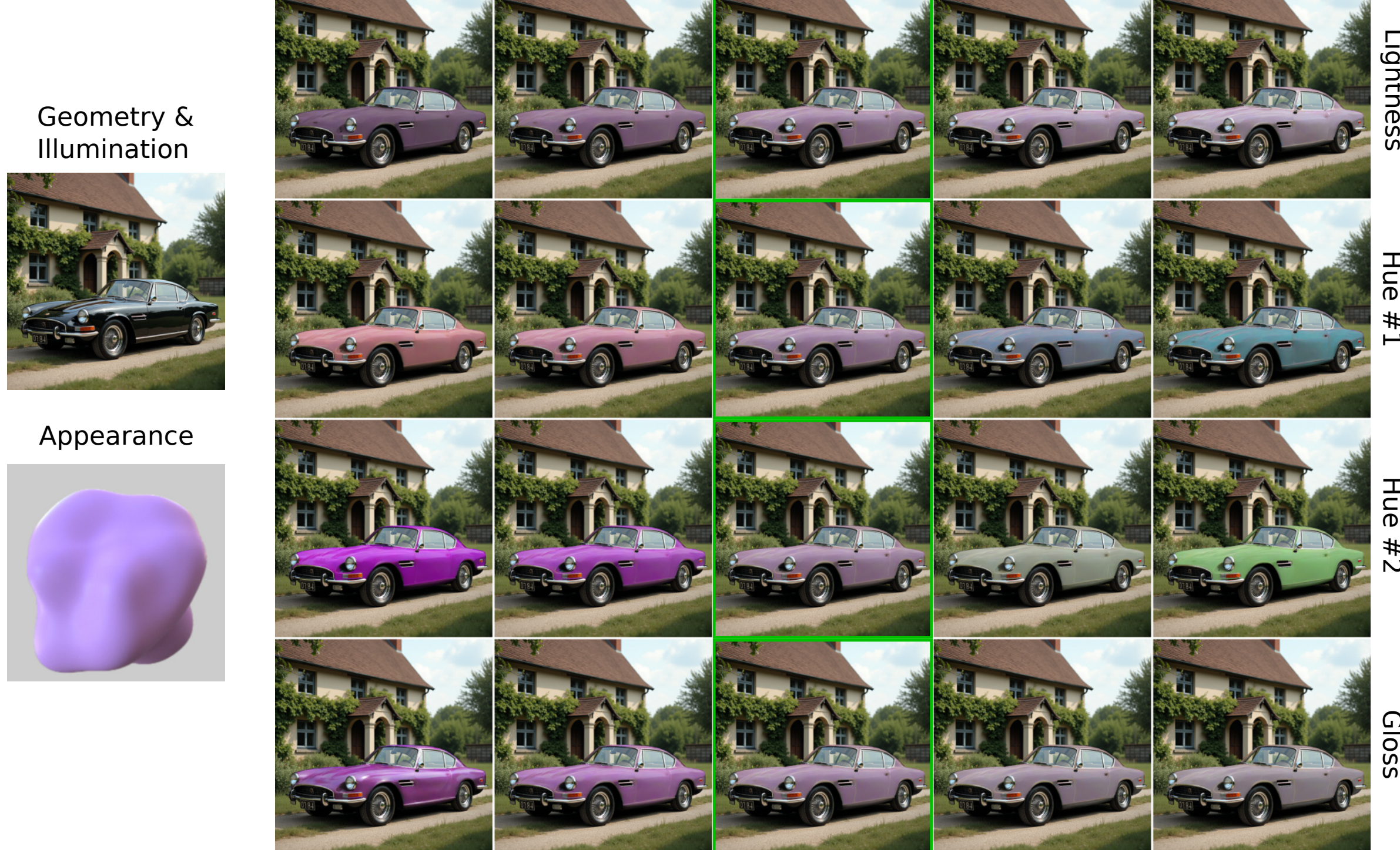

**Figure 21:** Posterior traversals *plot generated using* ***car*** *and* ***blob*** *as geometry and appearance references. Transfer marked in green.*

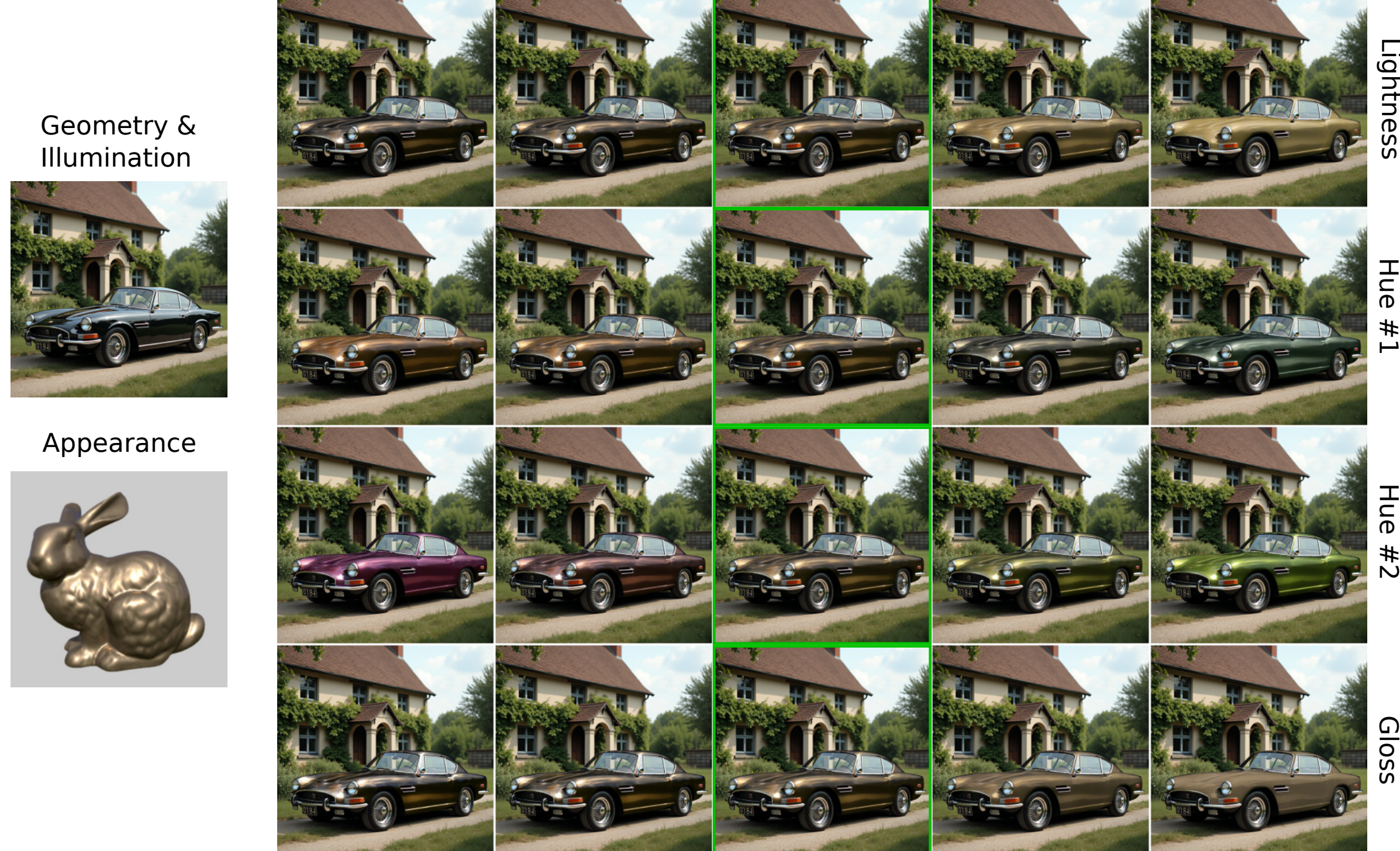

**Figure 22:** Posterior traversals *plot generated using* ***car*** *and* ***bunny*** *as geometry and appearance references. Transfer marked in green.*

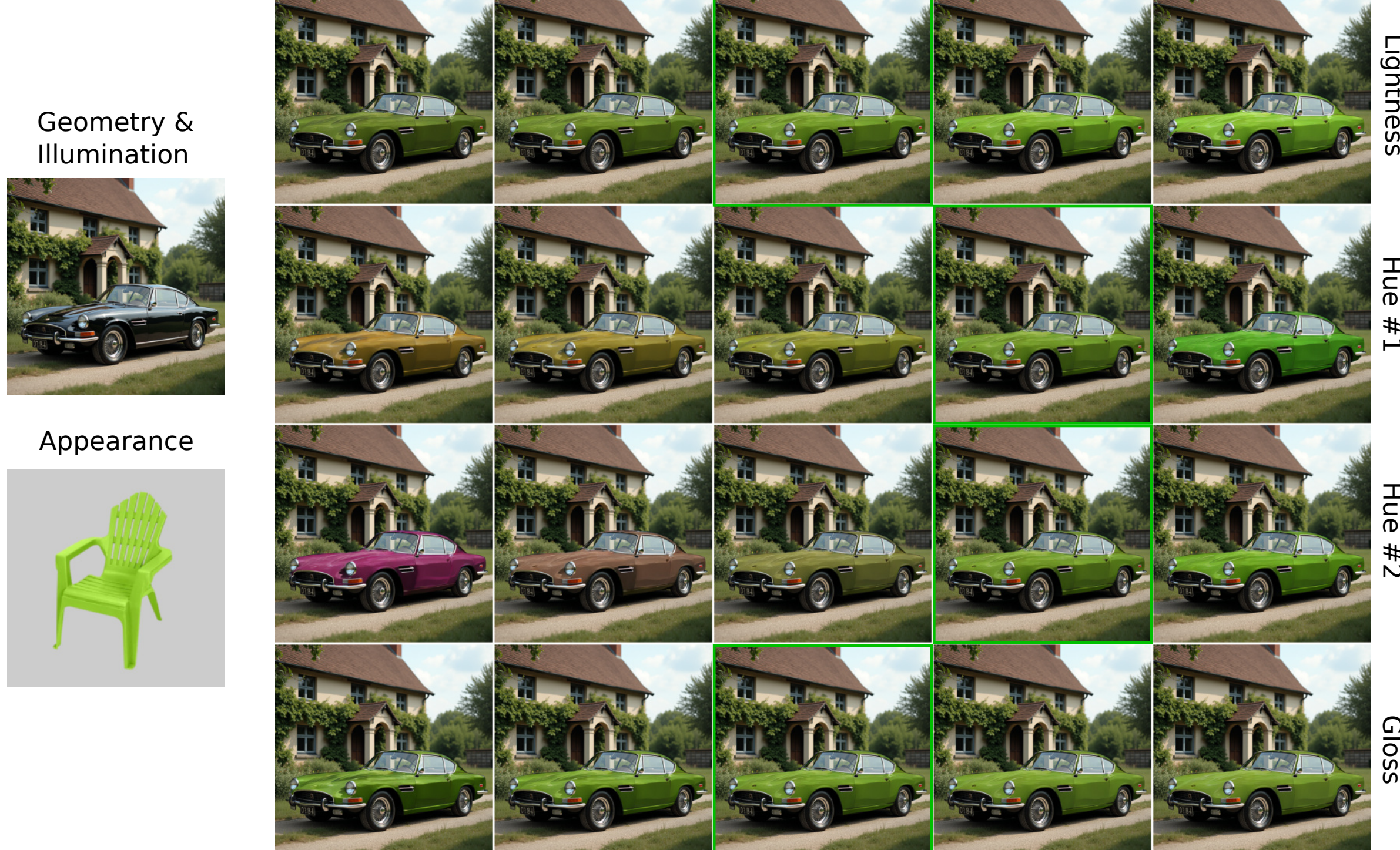

**Figure 23:** Posterior traversals *plot generated using* ***car*** *and* ***chair*** *as geometry and appearance references. Transfer marked in green.*

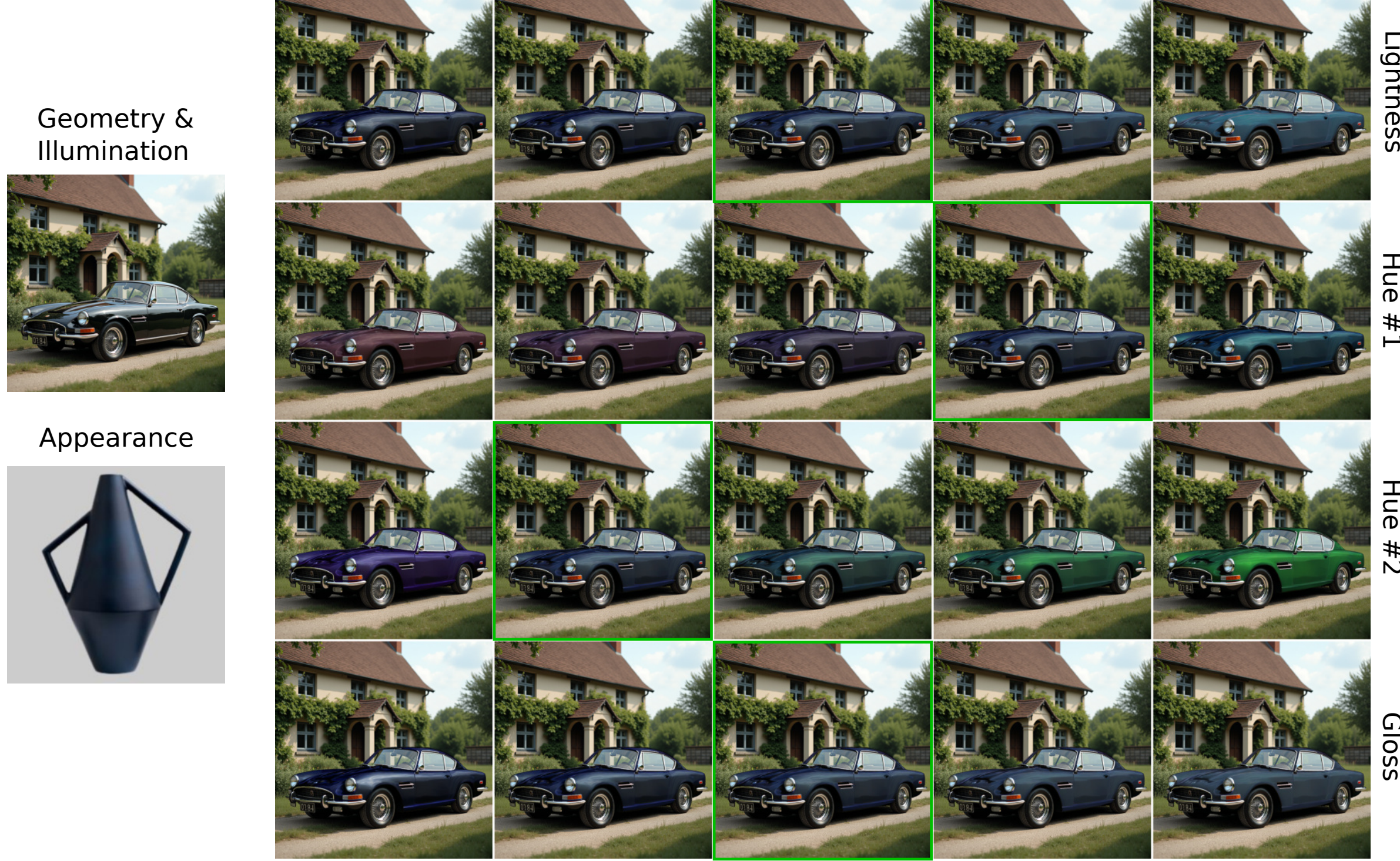

**Figure 24:** Posterior traversals *plot generated using* ***car*** *and* ***jar*** *as geometry and appearance references. Transfer marked in green.*

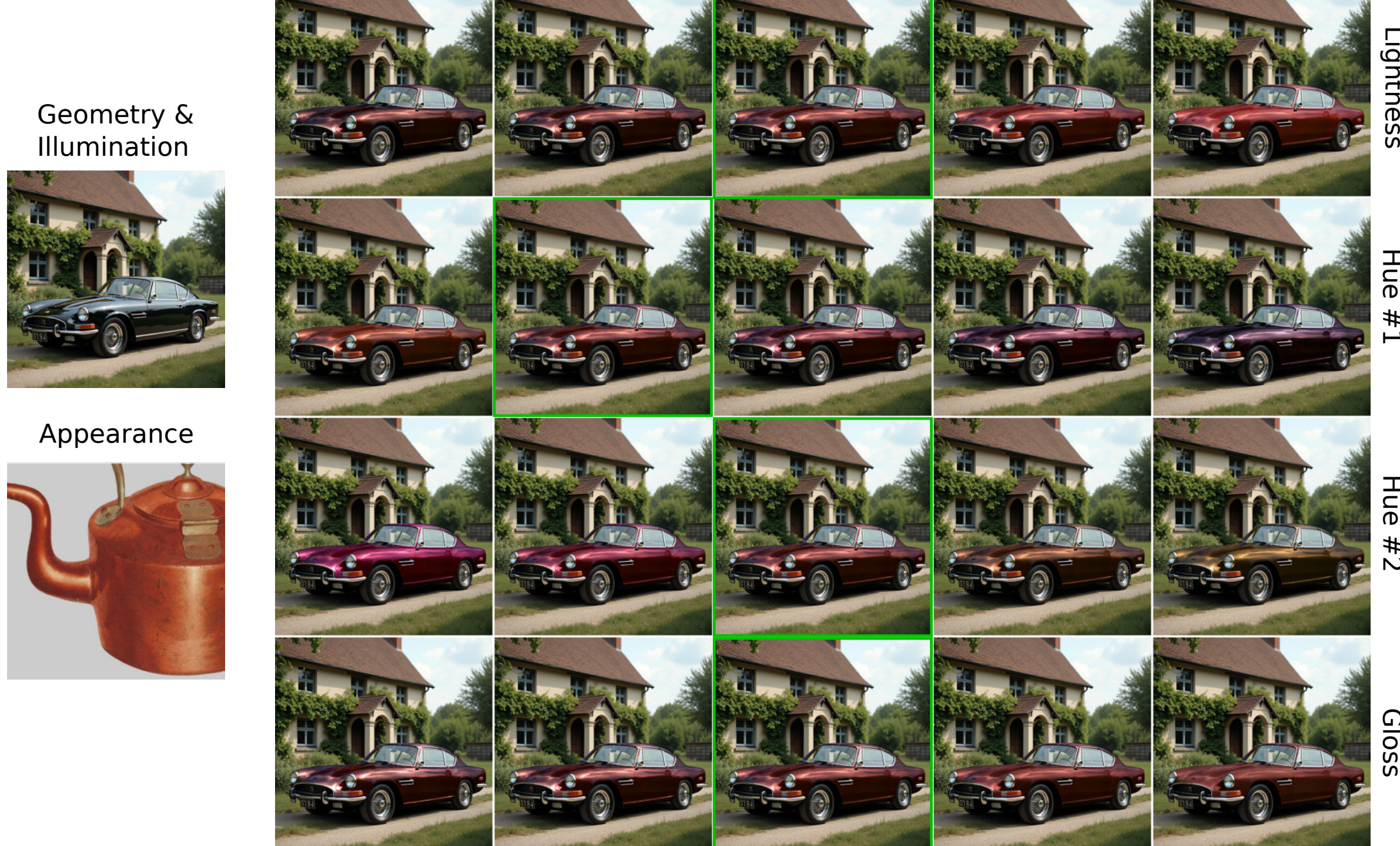

**Figure 25:** Posterior traversals *plot generated using* ***car*** *and* ***teapot*** *as geometry and appearance references. Transfer marked in green.*

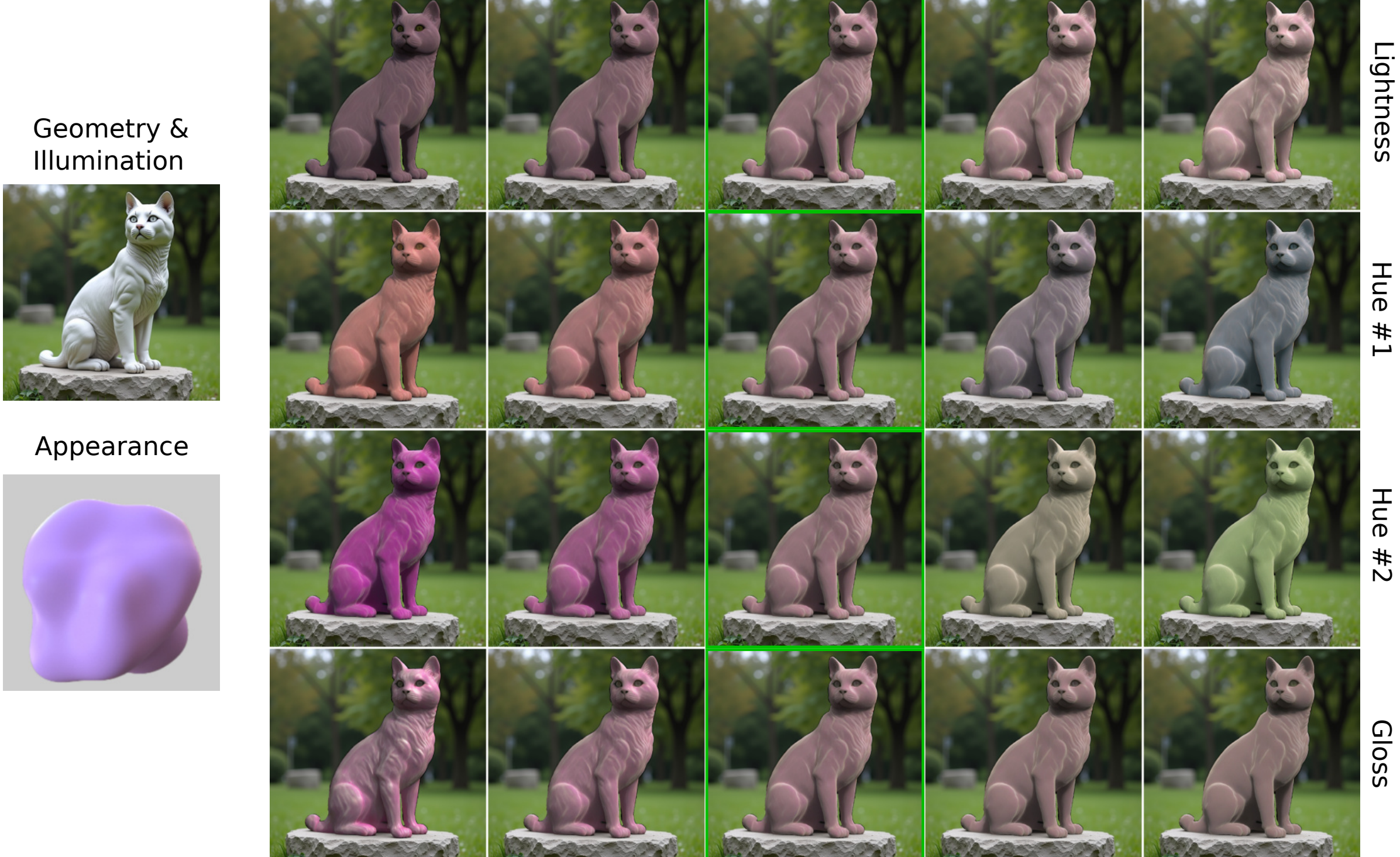

**Figure 26:** Posterior traversals *plot generated using* ***cat*** *and* ***blob*** *as geometry and appearance references. Transfer marked in green.*

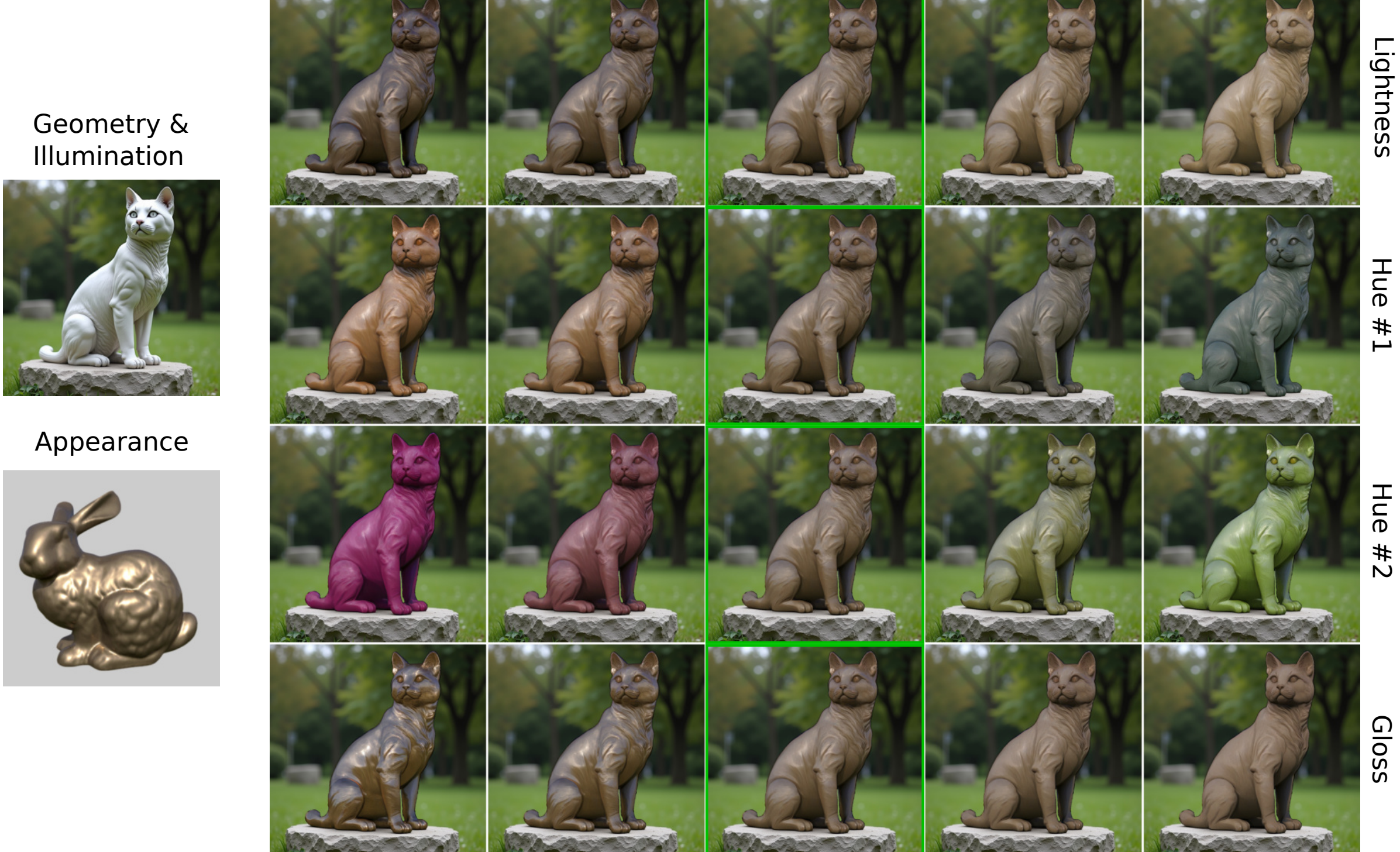

**Figure 27:** Posterior traversals *plot generated using **cat** and **bunny** as geometry and appearance references. Transfer marked in green.*

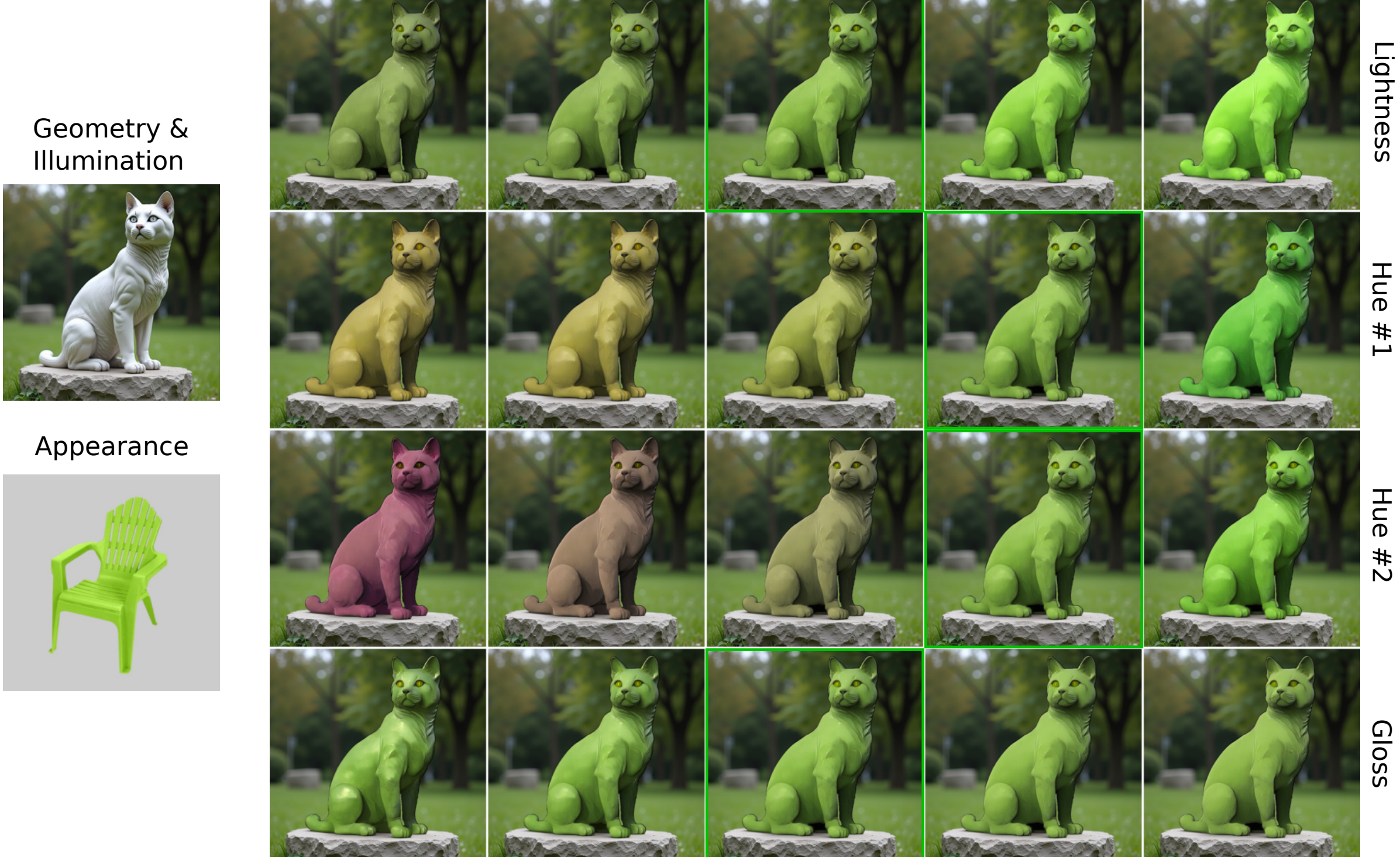

**Figure 28:** Posterior traversals *plot generated using **cat** and **chair** as geometry and appearance references. Transfer marked in green.*

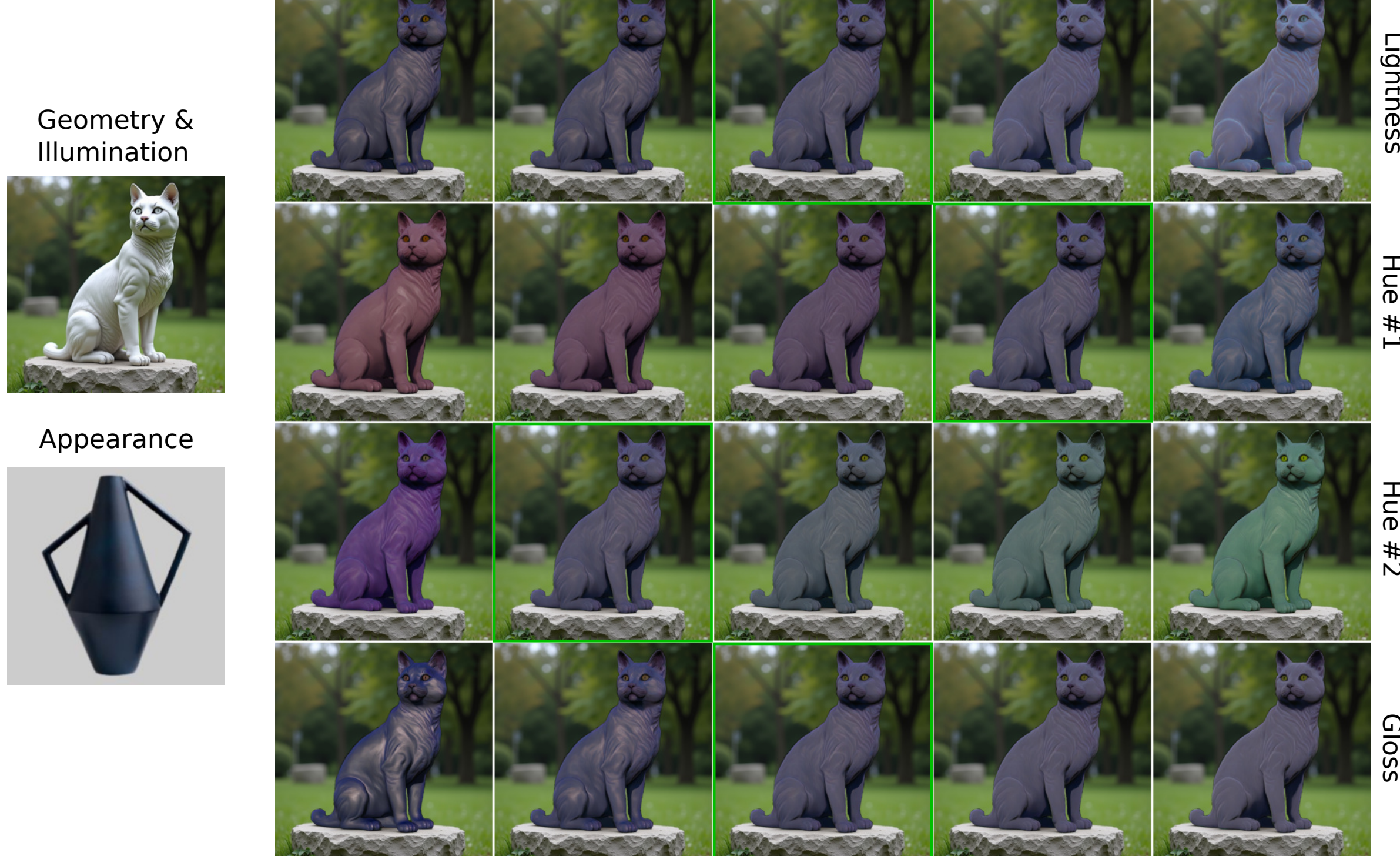

**Figure 29:** Posterior traversals *plot generated using* ***cat*** *and* ***jar*** *as geometry and appearance references. Transfer marked in green.*

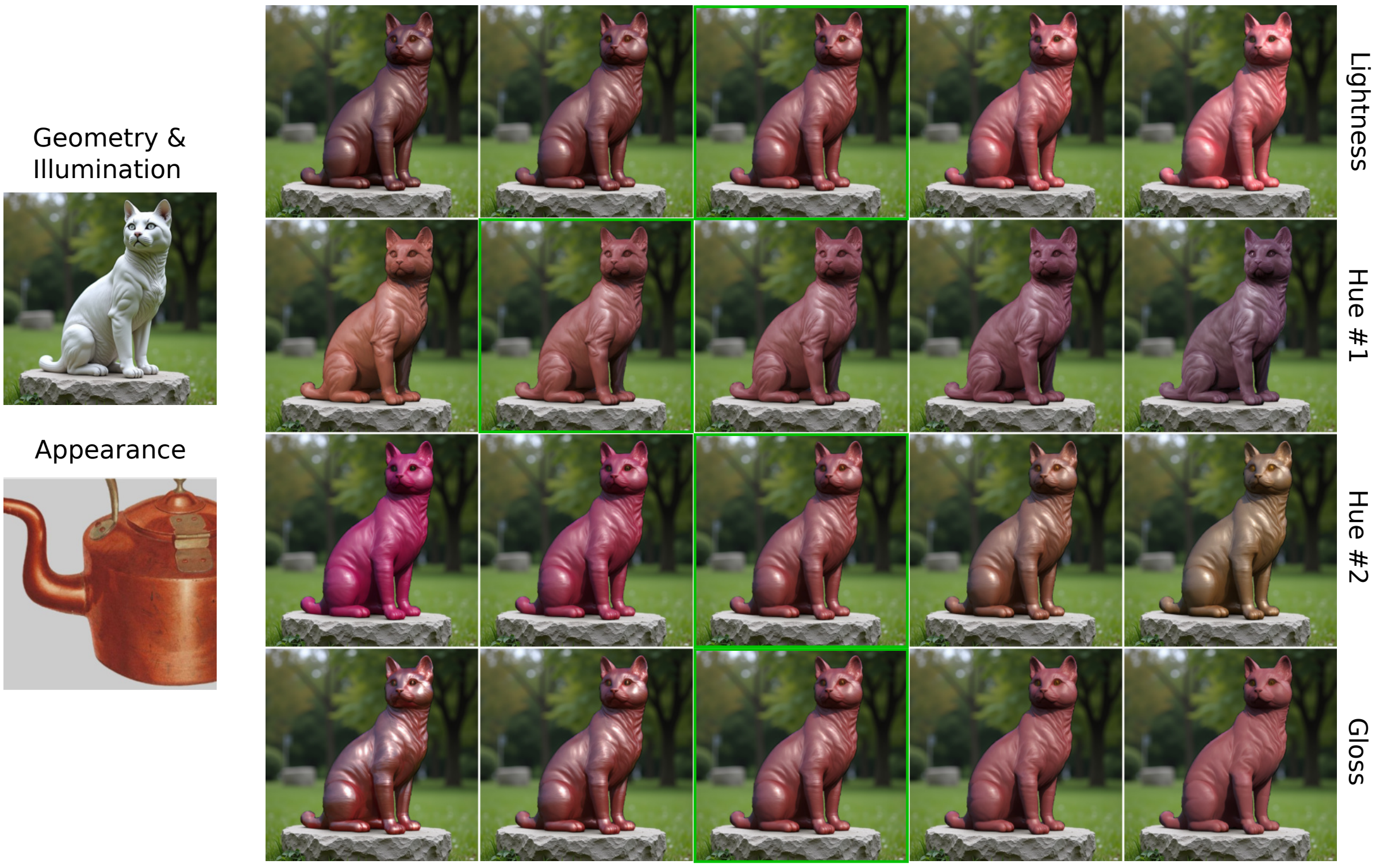

**Figure 30:** Posterior traversals *plot generated using* ***cat*** *and* ***teapot*** *as geometry and appearance references. Transfer marked in green.*

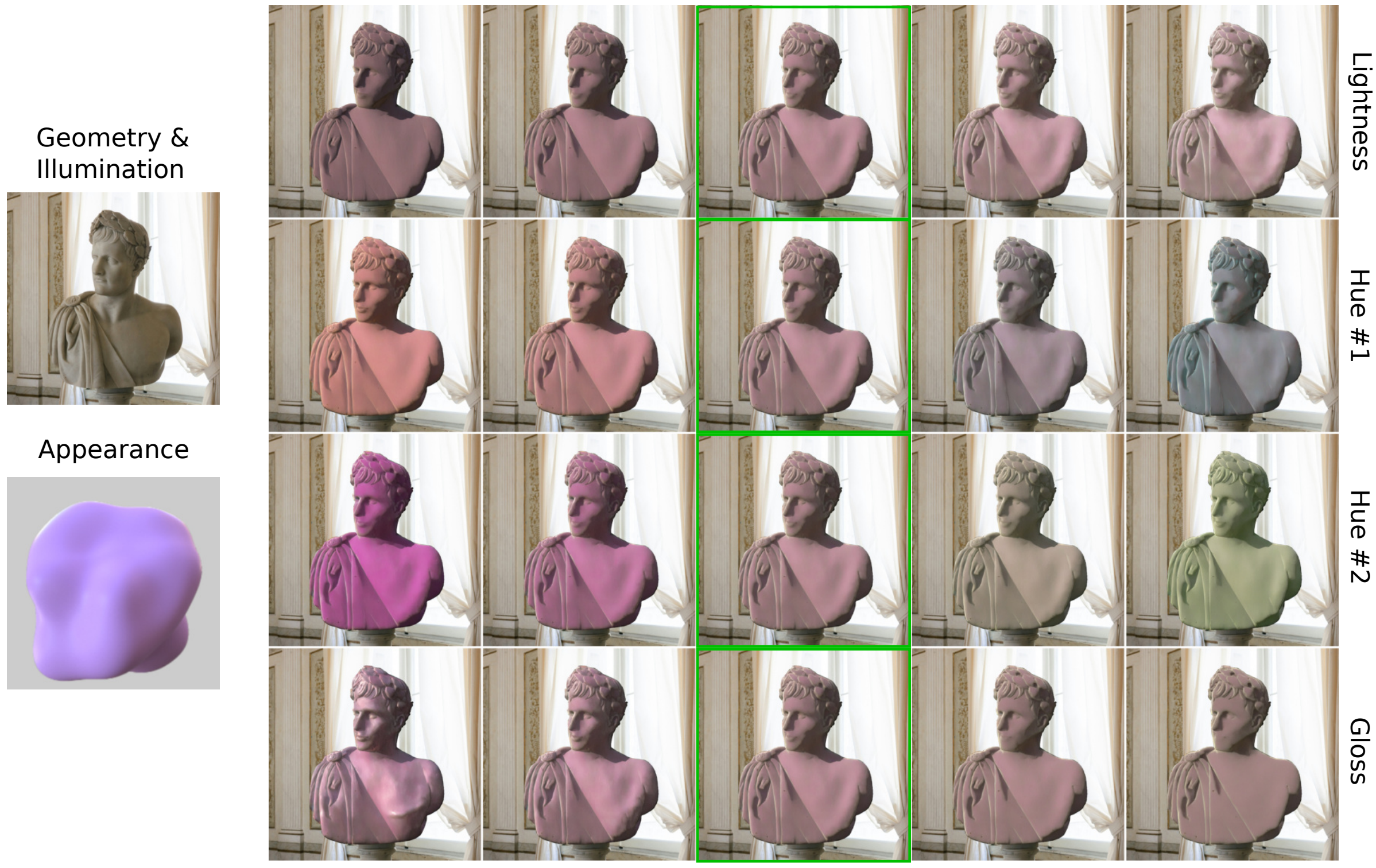

**Figure 31:** Posterior traversals *plot generated using* ***napoleon*** *and* ***blob*** *as geometry and appearance references. Transfer marked in green.*

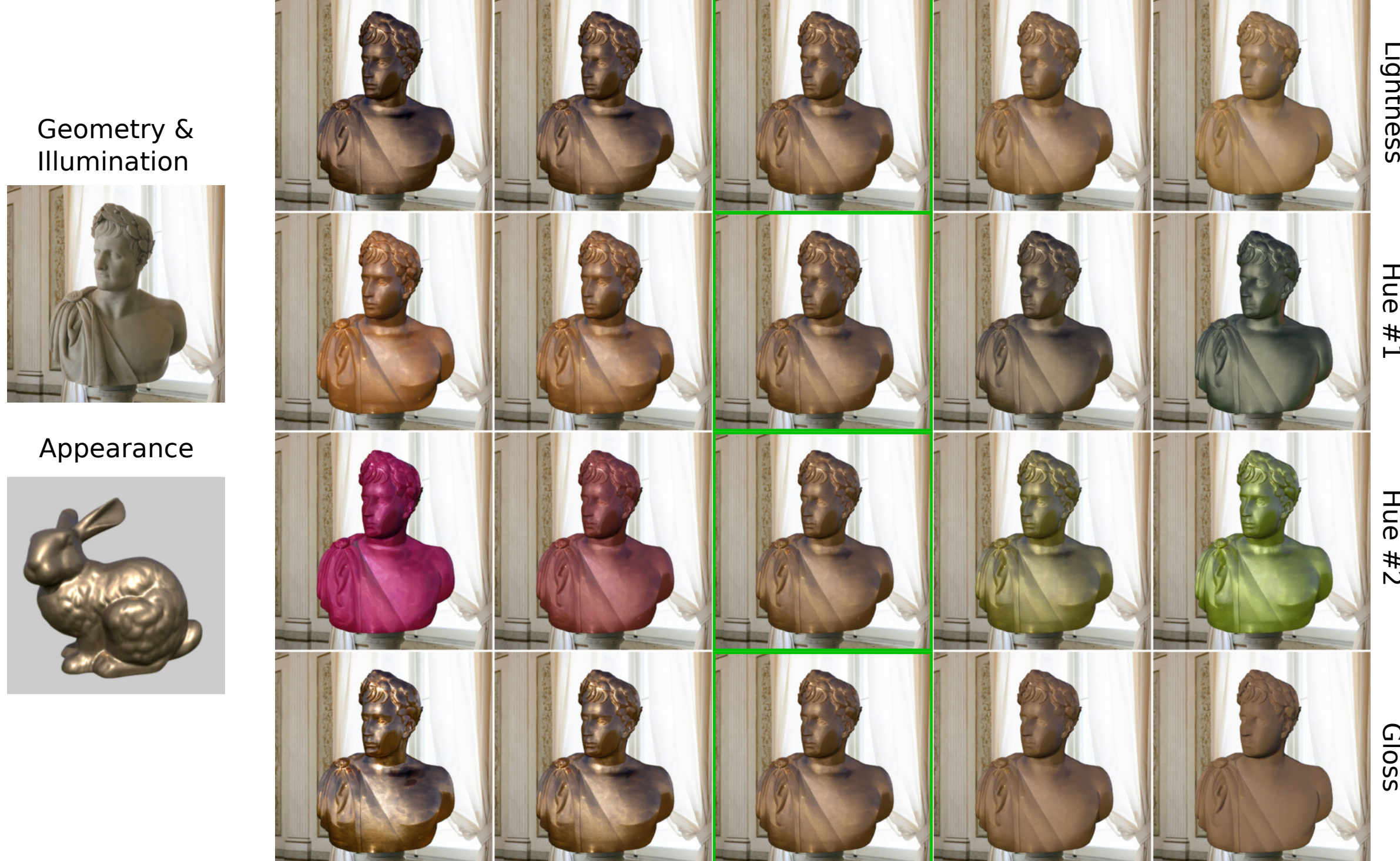

**Figure 32:** Posterior traversals *plot generated using* ***napoleon*** *and* ***bunny*** *as geometry and appearance references. Transfer marked in green.*

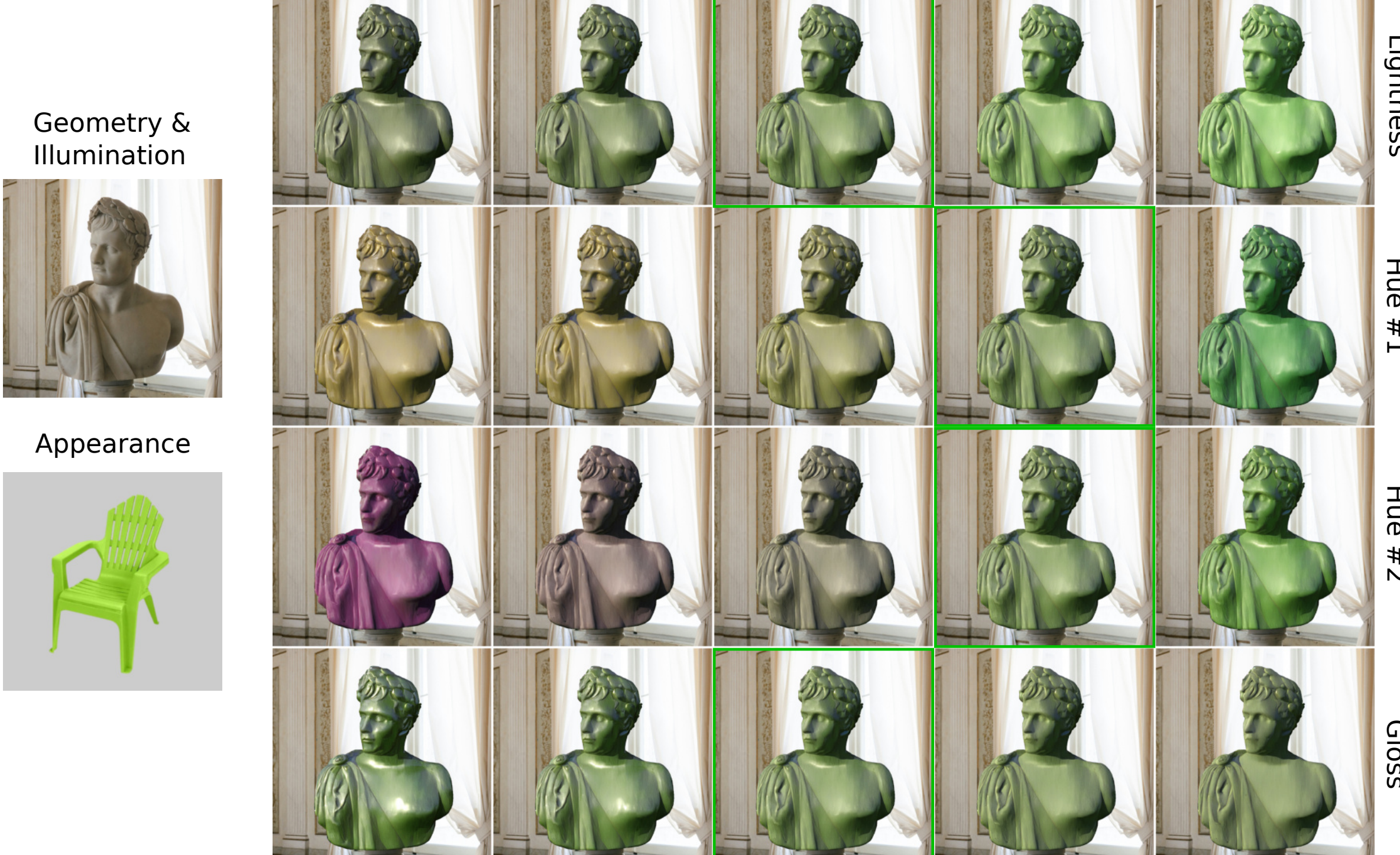

**Figure 33:** Posterior traversals *plot generated using **napoleon** and **chair** as geometry and appearance references. Transfer marked in green.*

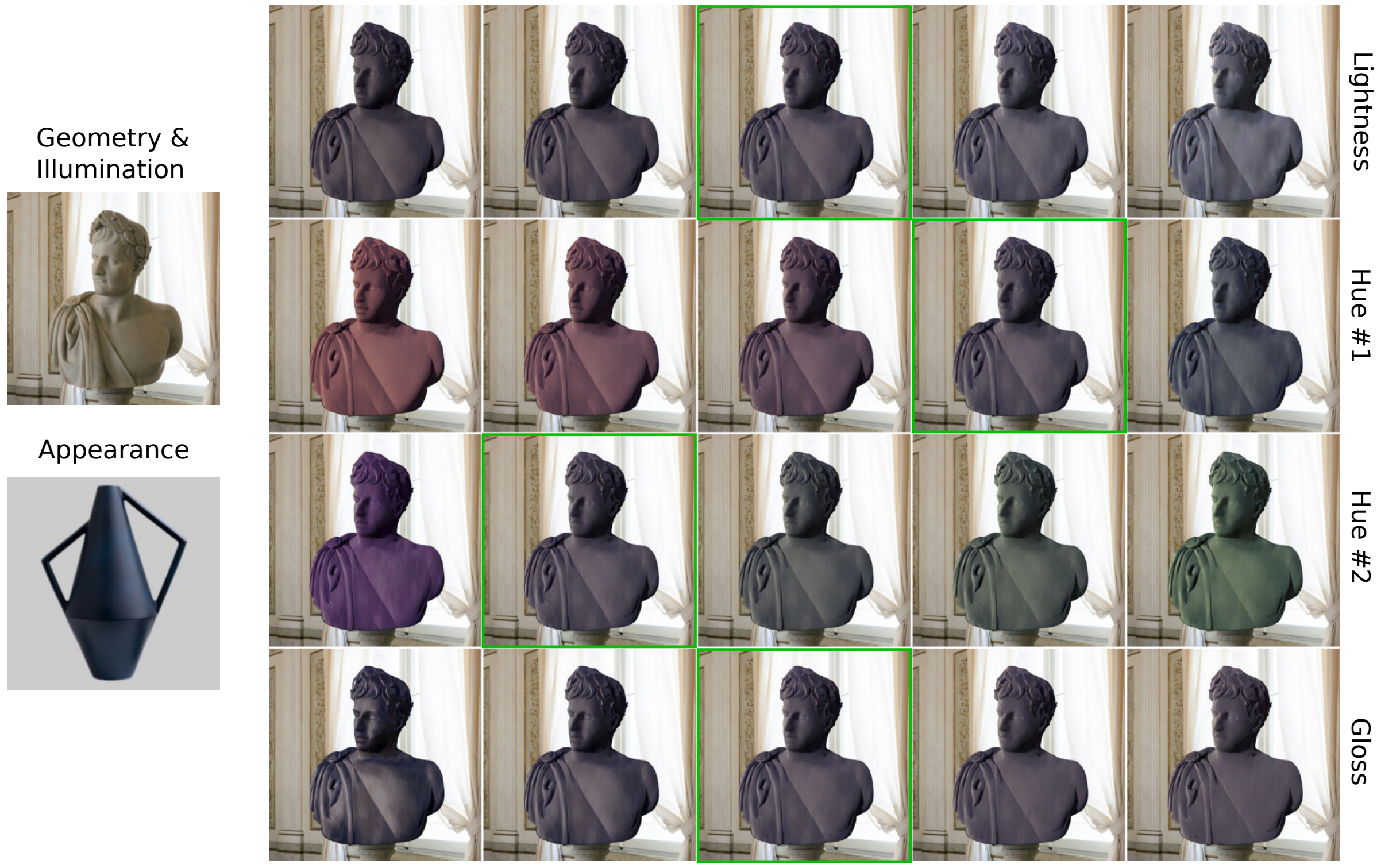

**Figure 34:** Posterior traversals *plot generated using **napoleon** and **jar** as geometry and appearance references. Transfer marked in green.*

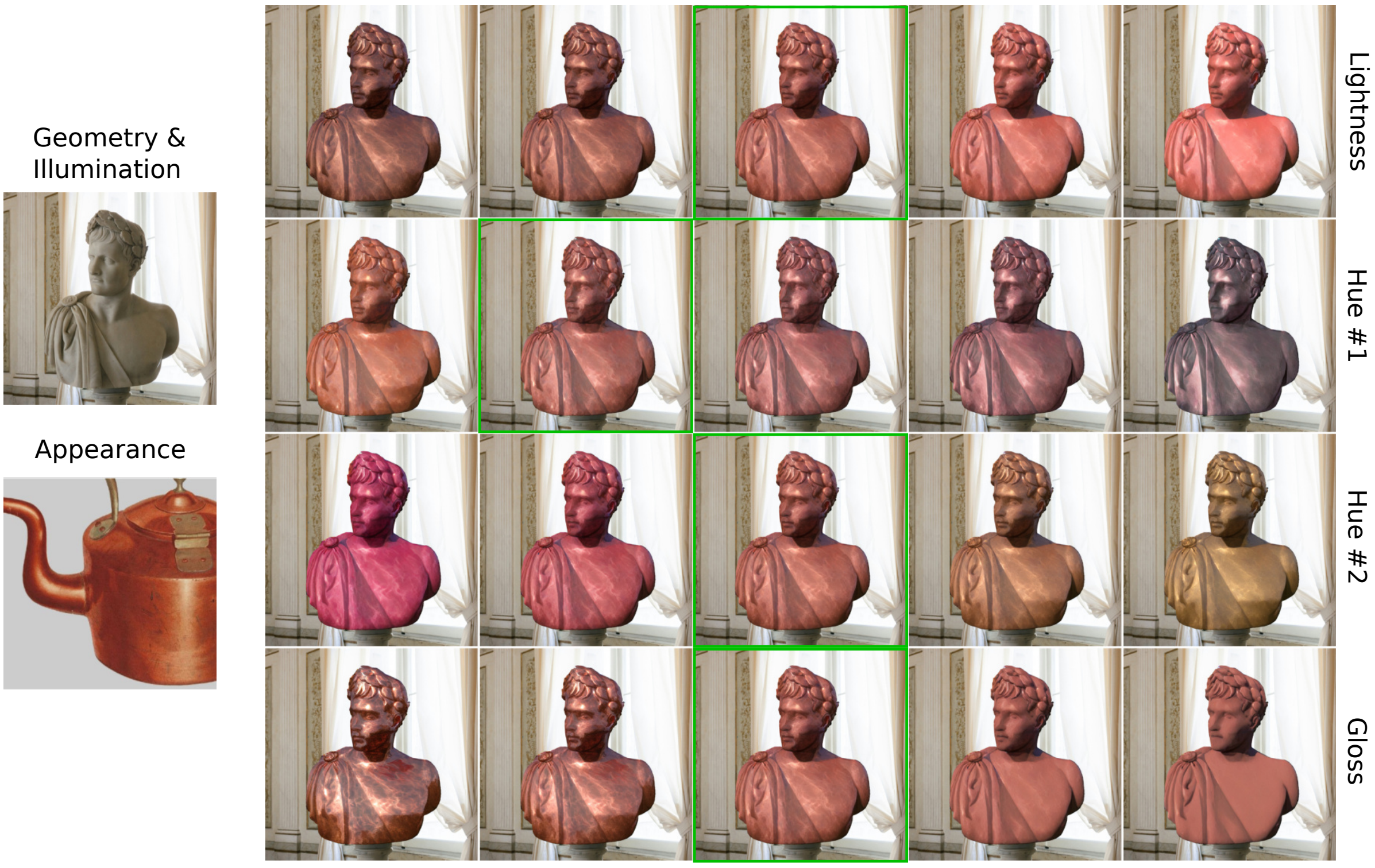

**Figure 35:** Posterior traversals *plot generated using* ***napoleon*** *and* ***teapot*** *as geometry and appearance references. Transfer marked in green.*